\documentclass[fleqn,usenatbib]{mnras}

\usepackage{newtxtext,newtxmath}

\usepackage[T1]{fontenc}

\DeclareRobustCommand{\VAN}[3]{#2}
\let\VANthebibliography\thebibliography
\def\thebibliography{\DeclareRobustCommand{\VAN}[3]{##3}\VANthebibliography}

\usepackage{graphicx}	
\usepackage{amsmath}	
\usepackage{newtxtext,newtxmath}
\usepackage[table]{xcolor}

\title[Spectral-timing study of kHz QPOs in $4U 1636-536$]{Spectral-timing analysis of the kilohertz quasi-periodic oscillations and constraints on the mass of the neutron star  in $4U 1636-536$ using {\em AstroSat} observations}

\author[Chattopadhyay, Mandal \& Misra]{
Suchismito Chattopadhyay$^{1,2}$\thanks{E-mail: suchismitochattopadhyay@gmail.com},Soma Mandal$^{1}$\thanks{E-mail: soma2778.wbes@gmail.com},
Ranjeev Misra$^{2}$ 
\\
$^{1}$Department of Physics, Government Girls' General Degree College ,7 Mayurbhanj Road,Kolkata-700023, India\\
$^{2}$Inter-University Centre for Astronomy and Astrophysics, Post Bag 4, Ganeshkhind, Pune 411007, India\\
}

\date{Accepted XXX. Received YYY; in original form ZZZ}

\pubyear{\the\year{}}

\begin{document}
\label{firstpage}
\pagerange{\pageref{firstpage}--\pageref{lastpage}}
\maketitle

\begin{abstract}
	
Kilohertz quasi-periodic oscillations (kHz QPOs) are believed to originate from the orbital timescales of the inner accretion flow, reflecting the dynamics of the innermost disk regions under strong gravitational forces. Despite numerous radiative and geometric models proposed so far, a comprehensive explanation of the observed properties of these variability components remains elusive. This study systematically examines kHz QPOs, their variability, and their connection to spectral properties in $4U 1636-536$ using {\em AstroSat} data. Our analysis tracks the source’s transition from hard to soft states in the hardness-intensity diagram. Broad spectral analysis (0.7–25 keV) using SXT and LAXPC data indicates a spectrum shaped by reflection from a thermal corona, with contributions from boundary layer emission and a soft disk component. We find significant changes in optical depth, blackbody temperature, and inner disk temperature that likely drive state transitions. Power density spectra reveal three variability types: a low frequency QPO (LFQPOs) ($\sim$ 30 Hz), and two simultaneous kHz QPOs. The LFQPOs and the upper kHz QPOs appear more prominently in soft spectral states. The presence of LFQPOs and twin kHz QPOs in soft spectral states enable us to estimate the neutron star mass at \(2.37 \pm 0.02\) $M_\odot$ using the relativistic precession model (RPM). Additionally, time-lag and root mean square (rms) analysis provide insights into the size of the corona and the radiative origin of these variability components.

\end{abstract}





\section{Introduction} \label{sec1}
Quasi-periodic oscillations (QPOs) are frequently observed and extensively studied features in the { power density spectra} (PDS) of low$-$mass X-ray binary systems (LMXBs) [{see} \citet{klitzing:wang2016brief} {for a review}]. PDS are typically modelled using multiple Lorentzian functions (e.g \citet{2003A&A...406..221S,2014AN....335..168W,2017MNRAS.468.2311M}), along with an additional constant to account for the Poisson noise level. The Lorentzian functions are characterized by three parameters, centroid frequency $\nu_0$, FWHM (full$-$width half maximum) which is related to the Quality factor ({$Q$}) \footnote{{$Q$} is defined as the ratio of the centroid frequency to the FWHM.} of the QPO and normalization, {which is associated with the root mean squared amplitude of the QPO \citep{10.1111/j.1365-2966.2006.10571.x,10.1093/mnras/stz2463}.} This fitting approach is independent of both the sources and the production mechanisms of QPOs, proving it particularly useful for comparing variability patterns across different systems \citep{klitzing:nowak2000there, klitzing:belloni2002unified, 10.1093/mnras/stac1071}.

{ In neutron star low-mass X-ray binaries (NSLMXBs), the centroid frequency of the QPOs ranges from very low frequencies ($\sim$ mHz) \citep{10.1093/mnras/staa3224,10.1093/mnras/stad949} to very high frequencies (up to about 1300 Hz) \citep{2009MNRAS.399.1901B}, displaying a wide range of variability}. They are sub$-$classified majorly into two categories depending on their frequency and characteristics. {The QPOs with low frequency (5$-$60 Hz, {$Q \geq 2$}) are called low frequency QPOs (LFQPOs) \citep{Wijnands_1998, vanderKlis_2006} and the higher frequencies ranging mostly from 400$-$1300 Hz are classified as  High$-$Frequency QPOs or kHz QPOs \citep{VanDerKlis_1997, klitzing:peille2015spectral} and see \citet{klitzing:wang2016brief} for a review}.

These kHz oscillations often appear in pairs, with the higher centroid frequencies referred to as the upper kHz QPOs and the lower ones as the lower kHz QPOs \citep{1996ApJ...469L...1V,1997ApJ...485L..37M,1998ApJ...505L..23M,2000astro.ph..8358J, 2006MNRAS.366.1373Z, klitzing:wang2016brief,klitzing:M_ndez_2020}. Although several theoretical models have been proposed to explain their dynamical origin, such as the Magnetospheric beat frequency model \citep{Alpar1985}, the sonic$-$point beat frequency model \citep{1998ApJ...508..791M}, the relativistic precision model (RPM) \citep{1998ApJ...492L..59S}, the disk oscillation model \citep{Titarchuk_1998}, the Magneto hydrodynamical (MHD) Alfven Wave Oscillation Model \citep{2004A&A...423..401Z}), a definitive explanation remains elusive. The time scale of these kinds of variabilities {is} comparable to the dynamical time scale of the inner disk region, which allows us to study the dynamics of the  accreted material rotating in extreme gravity near the neutron star and map the curved space time around the neutron star \citep{klitzing:1996ApJ...469L...9S,klitzing:wang2016brief, klitzing:M_ndez_2020}. Around a neutron star (NS) the orbital frequency of in$-$falling matter can be expressed using the following equation,

\begin{equation}
\nu_{\text{orb}} = \sqrt{\left(\frac{GM}{4\pi^2 r^3}\right)}\quad \approx \quad 1200 \left(\frac{r_{\text{orb}}}{15 \text{km}}\right)^{-3/2} M_{1.4\odot}^{1/2} \quad \text{Hz}
\label{E1}
\end{equation}

This aligns with the kHz QPO frequencies observed in NS LMXBs, linking these frequencies to the orbital motion of material near neutron stars ($M_{1.4\odot}$) \citep{Zhang_1997} and see \citet{2006csxs.book...39V,klitzing:wang2016brief} for a review. Later, the sonic-point beat frequency model \citep{1998ApJ...508..791M} was proposed, where the upper kHz QPO is ascribed to the Keplerian motion of the orbiting material and the lower one has been ascribed to the difference (beat frequency) between the Keplerian frequency and the spin frequency of the NS.
This model explains kHz QPO frequencies and their consistency across Z and atoll sources, assuming that the moderate magnetic fields do not disrupt the formation of a sonic point. It also explains the high coherence, large amplitudes, and increasing amplitude with photon energy. As more kHz QPO data became available, it was observed that the frequency separation between the twin QPOs is sometimes significantly smaller than the neutron star spin frequency \citep{1997ApJ...481L..97V,1998ApJ...505L..23M,2003astro.ph..8179L}. This discrepancy suggests that the beat-frequency model does not fully align with all observations.


On the other hand, based on the idea that in the inner disk region the frame dragging and compactness will lead to the observable relativistic effects, the Relativistic Precessional Model (RPM)  \citep{1998ApJ...492L..59S,1999ApL&C..38...57S,1999PhRvL..82...17S} has been put forward. In General relativity, the particle orbits are not closed (i.e. they will not come to the same point after one revolution), as the azimuthal, radial, and vertical motions differ from each other. Due to this difference in the azimuthal frequency ($\nu_{\phi}$), radial frequency ($\nu_{r}$) and vertical frequency ($\nu_{\theta}$) eccentric orbits precess at the Periastron precession frequency \(\nu_{\text{peri}} = \nu_{\phi} - \nu_{r}\), while orbits inclined relative to the equatorial plane of a rotating central mass experience nodal precession at the frequency 
\(\nu_{\text{nodal}} = \nu_{\phi} - \nu_{\theta}\) \citep{1972ApJ...178..347B,10.1046/j.1365-8711.1999.02307.x}. In this model, the frequency separation \(\Delta \nu\) is predicted to decrease with increasing \(\nu_\phi\) (as observed; \citet{1999PhRvL..82...17S,1999ApL&C..38...57S}) and with sufficient decreases in \(\nu_\phi\) as well. Moreover, The nodal precession frequency \(\nu_{\text{nodal}}\) incorporates a structure$-$dependent term \(I/M\) \footnote{I is the Moment of inertial in units of $10^{45}$ gm$-$cm$^2$ (I$_{45}$).}. Therefore, measuring a kHz QPO peak linked to stable Keplerian motion allows us to constrain the neutron star mass \(M\) and radius \(R\).

Since these variabilities are expected to originate from the inner disk region close to the NS, one may retrieve information about the properties of the accretion flow and the corresponding environment through studying different properties related to these kind of variabilities. For instance, \citet{klitzing:barret2013soft} shows that the lower kHz QPO frequency decreases with the increasing inner disk radius, which is expected from Equation \ref{E1}. {For 4U 1636$-$536, the spectral photon index ($\Gamma$) initially increases and then decreases as the frequency of the upper kHz QPO increases \citep{10.1093/mnras/stx1686}. In contrast, for 4U 0614+091, $\Gamma$ increases monotonically with QPO frequency \citep{1998ApJ...497L..93K}.} Meanwhile, electron temperature ($k_BT_e$) shows an opposite decreasing variation to the increasing frequency for both the lower and the upper kHz QPO in 4U 1636$-$536 \citep{10.1093/mnras/stx1686}. These findings suggest a deeper connection between the origin of QPOs and spectral parameters, which could be explored further through time lag and rms variation studies. 

Time lags refer to the delay between photon arrival times in the hard and soft energy bands. Lower kHz QPOs typically exhibit soft lags $(\sim\mu s)$, where photons in the soft energy band arrive later than those in the hard band \citep{klitzing:barret2013soft,klitzing:de2013time,klitzing:kumar2014energy,klitzing:2018ApJ...860..167T,klitzing:10.1093/mnras/stab1905} while an opposite trend can be seen in the case of upper kHz QPOs \citep{klitzing:de2013time, klitzing:peille2015spectral, klitzing:10.1093/mnras/stab1905}. The short time lags of around $ 50~\mu s$  suggest that they are linked to the Compton scattering timescale within a medium of approximately 10 km \citep{klitzing:lee2001compton, klitzing:kumar2014energy, klitzing:kumar2016constraining}. In general, one expects that Compton scattering should produce hard lags \citep{klitzing:wijers1987energy, klitzing:lee1998comptonization}, but soft lags can also be produced, and such a model can explain the observations, if there is a fraction of the Comptonized photons impinging back into the soft photon source \citep{klitzing:lee2001compton, klitzing:kumar2014energy, klitzing:kumar2016constraining,10.1093/mnras/stz3502,2022MNRAS.515.2099B}.  Further, \citet{2022A&A...662A.118M} modified the model by incorporating a finite radiative cooling timescale for the corona. By introducing a time-dependent evolution of the coronal temperature in response to variations in the heating and radiative cooling rates they have shown that, under certain parameter choices, the system can display strong oscillatory behaviour. Since, typically for  black hole systems, the coronal radiative cooling timescale is less than $0.1$ milliseconds, they have associated this timescale with LFQPOs , considering  a large coronal size and accretion rates corresponding to a luminosity of $\sim 10^{35}\ \mathrm{ergs\,s^{-1}}$, significantly less than the commonly observed luminosities of $\sim 10^{38}\ \mathrm{ergs\,s^{-1}}$. Moreover, this modification has been shown to account for the kHz QPO timescale by requiring a mass of $\leq 1\,M_{\odot}$, which is nevertheless below the established lower limit for neutron star masses of about $1.4\,M_{\odot}$.

4U 1636$-$536 is a persistent atoll source, first discovered by \citet{10.1093/mnras/169.1.7} with a short orbital period of 3.79 Hrs \citep{Giles_2002} evaluated from the photometry of optical light curve.  Assuming the faint radius$-$expansion bursts reach the Eddington limit for hydrogen$-$rich material (X $\approx$ 0.7) and the brighter bursts reach the limit for pure helium (X = 0), \citet{Galloway_2006} estimate the distance to 4U 1636$-$536 to be approximately $6.0 \pm 0.5 \, \text{kpc}$ for a canonical neutron star with mass \(M_{NS} = 1.4 \, M_{\odot}\) and radius \(R_{NS} = 10 \, \text{km}\). For a neutron star with mass \(M_{NS} = 2 \, M_{\odot}\), the distance could be as much as $7.1 \, \text{kpc}$. \citet{10.1111/j.1365-2966.2006.11106.x} reports a mass function for 4U 1636–536 of $ f(M) = 0.76 \pm 0.47 \, M_{\odot} $ and a mass ratio of $
\frac{M_2}{M_{NS}} \approx 0.21 - 0.34,$ where \(M_2\) is the donor mass. They have also estimated the inclination angle to be between \(36^\circ\) and \(74^\circ\). The source exhibits frequent Type I bursts. { A burst oscillation frequency of 581 Hz has been observed, which is interpreted as the neutron star spin frequency \citep{Strohmayer_1998,10.1093/mnras/stad949}. \citet{1999ApJ...515L..77M} initially reported a weak and marginally significant signal at 290 Hz and suggested it as the fundamental, with 581 Hz as the first overtone. However, subsequent observations have not confirmed this interpretation, and the 581 Hz signal remains the widely accepted spin frequency .} Recent studies on Type I bursts using three distinct {\em AstroSat} observations by \citet{2021MNRAS.508.2123R} have further confirmed the existence of burst oscillation at $\sim$ 581 Hz. Moreover, twin kHz QPO has been observed in the PDS of this system using {\em Rossi X$-$ray Timing Explorer (RXTE)} where the lower kHz QPO frequency ranges between 644$-$769 Hz and the upper kHz QPO ranges between $\sim$ 900$-$1050 Hz (\citet{1996ApJ...473L.135Z,2002MNRAS.336L...1J,10.1111/j.1365-2966.2005.09214.x,10.1111/j.1365-2966.2007.11943.x,2011ApJ...726...74L} and references therein ).

In this work, we study the broadband spectral and temporal properties of 4U 1636$-$536 and their relation to the twin kHz QPOs observed in the PDS using 4 different {\em AstroSat} observations. The objective is to leverage the instruments' wide-band capabilities to constrain the energy spectra and subsequently quantify and model the source's temporal properties and to constrain the mass and radius of the system. Section \ref{sec2} outlines the observations and data reduction procedures, while Section \ref{sec3} focuses on the light curves and hardness ratios. The spectral analysis is detailed in Section \ref{sec4}, and Section \ref{sec5} covers the timing analysis and explores the energy-dependent temporal properties. Modelling the energy$-$dependent lag and rms using the spectral information has also been elaborated in Section \ref{sec5}. Measurement of the mass of the neutron star is described in Section \ref{sec6}.  Finally, Section \ref{sec7} summarises and discusses the results of our study.

\section{Observation and Data reduction} \label{sec2}

{\em AstroSat} has observed 4U 1636$-$536 on several occasions since 2016. To align with our scientific objectives, we initially screened the datasets based on the clear presence of kHz QPOs. Details of the final data sets that have been considered for further analysis have been presented in Table \ref{Table1}. 

To analyse the {Large Area X$-$ray Proportional Counter} (LAXPC) data, we utilised the {\tt LAXPCsoftware22Aug15}, which is publicly available on the official {AstroSat Science Support Cell (ASSC)} website\footnote{\url{http://astrosat-ssc.iucaa.in/laxpcData}}. We have merged all orbit files to create a single level 2 event file, from which further scientific products have been derived using good time intervals, adhering to the standard procedures outlined on the ASCC webpage\footnote{\url{http://astrosat-ssc.iucaa.in/uploads/threadsPageNew_SXT.html}}. It is crucial to mention that all LAXPC {Proportional Counter Units} (PCUs)—specifically, PCU 10, PCU 20, and PCU 30—were included in the analysis. However, for the spectral analysis, only PCU 20 has been considered instead of {PCU 10 or 30} as PCU 10 has a higher background count as one of its veto anodes is not functioning properly \citep{klitzing:2017ApJS..231...10A}, and PCU 30 has some gain-related issue and stopped operating on March 8, 2018, due to gas leakage \cite{klitzing:Antia_2021}.

{\tt SXTPIPELINE}\footnote{\url{https://www.tifr.res.in/~astrosat\_sxt/sxtpipeline.html}} has been used to process the SXT level1 data to produce the level2 data. Level2 data from different orbits have been merged using the Julia$-$based module {\tt SXTEVTMERGERTOOL} \footnote {\url{http://astrosat-ssc.iucaa.in/sxtData}} to produce the merged event file, which is further analyzed using Heasoft v6.29 routine {\tt XSELECT} to get the advanced scientific products. The average SXT count rate comes out to 10$-$15 counts/sec using a region of 480 arc$-$second {radius} which is well below the 40 counts/sec, hence no pile$-$up correction has been initiated. For spectral analysis, we only use strictly simultaneous SXT data alongside the LAXPC data. It is to be noted further that, all the light curves exhibit clear presence of Type 1 bursts, which have been excluded from the analysis. Only the persistent segments of the light curves are analyzed.   

\begin{table}
	\centering
	\begin{tabular}{c c c c}
		\hline
		{\bf Observation} & {\bf Obs.Date} & {\bf Exp. Time} & {\bf Name} \\
		{\bf ID} & & (sxt / laxpc)& \\
		\hline
		\hline 
		G05\_195T01\_9000000530 & 02/07/2016 & 15 / 39 (ksec) & Obs 1\\
		\hline
		G05\_195T01\_9000000598 & 13/08/2016 & 14 / 38 (ksec) & Obs 2\\
		\hline
		G05\_002T01\_9000000616 & 20/08/2016 & 14 / 33 (ksec) & Obs 3\\
		\hline
		A04\_055T01\_9000001574 & 02/10/2017 & 14 / 52 (ksec) & Obs 4 \\
		\hline
	\end{tabular}
	\caption{The details of the Observations that has been used in this work for the further analysis. The observations are screened on the basis of the presence of the kHz QPO. The Observational data are available on the {\em AstroSat } data web page (\url{ https://astrobrowse.issdc.gov.in/astro\_archive/archive}).}	
	\label{Table1}
\end{table}

\section{Evolution of the Source in Hardness Intensity Diagram (HID)} \label{sec3}

\begin{figure}
	\centering
	\begin{tabular}{c}
		\includegraphics[scale=0.4]{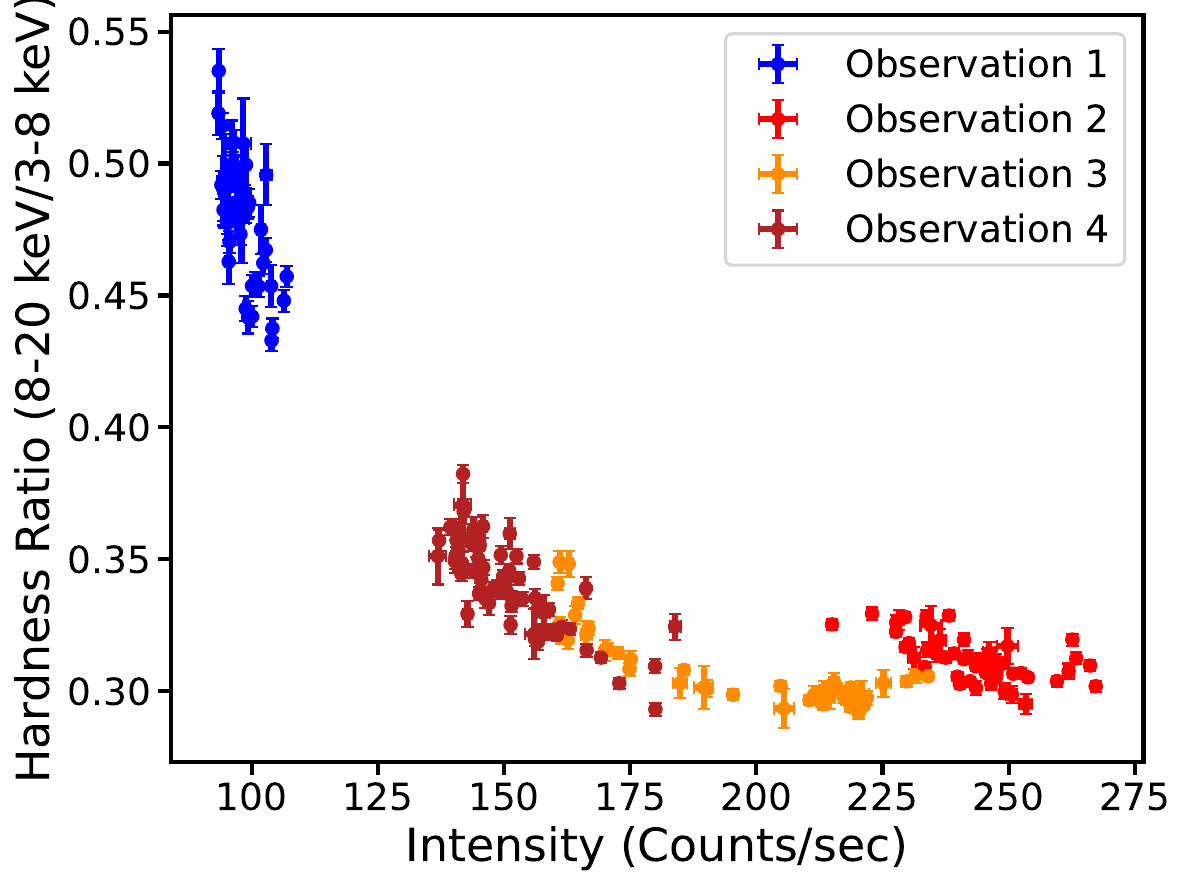}
	\end{tabular}
	\caption{Hardness vs Intensity diagram using LAXPC PCU 20 to see the evolution of {4U 1636$-$536}. The bin time is 1024 sec. }
	\label{Fig00}
\end{figure}

Figure \ref{Fig00} illustrates the source trajectory on the hardness$-$intensity diagram (HID). We used the background$-$subtracted PCU 20 light curve with a bin size of 1024 seconds to generate the HID. The hardness ratio is defined as the count rate ratio between the 8$-$20 keV and 3$-$8 keV energy bands, while the intensity is the sum of the counts in both energy bands. The source resides predominantly in the softer region, except for Obs 1. As the system transits to a softer state, the count rates increase as expected.

\section{spectral analysis} \label{sec4}
\label{sec:4}

We have conducted a joint spectral analysis using strictly simultaneous data from LAXPC and SXT in the persistent light curve segments covering a broad energy range of 0.7$-$25 keV. For the LAXPC spectrum, we focused on the 3$-$25 keV range, as the background count rate becomes dominant above 25 keV, while the SXT spectrum spans over 0.7$-$7 keV. All the spectra are further optimally grouped using {\tt ftgrouppha} and a 3 \% systematic has been added to all spectral fitting. Additionally, a gain correction with a slope fixed at 1 has been applied to the SXT spectra throughout the analysis to take into account the linear variation of gain with energy.

\begin{figure}
	\centering
	\begin{tabular}{c}
		\includegraphics[scale=0.4]{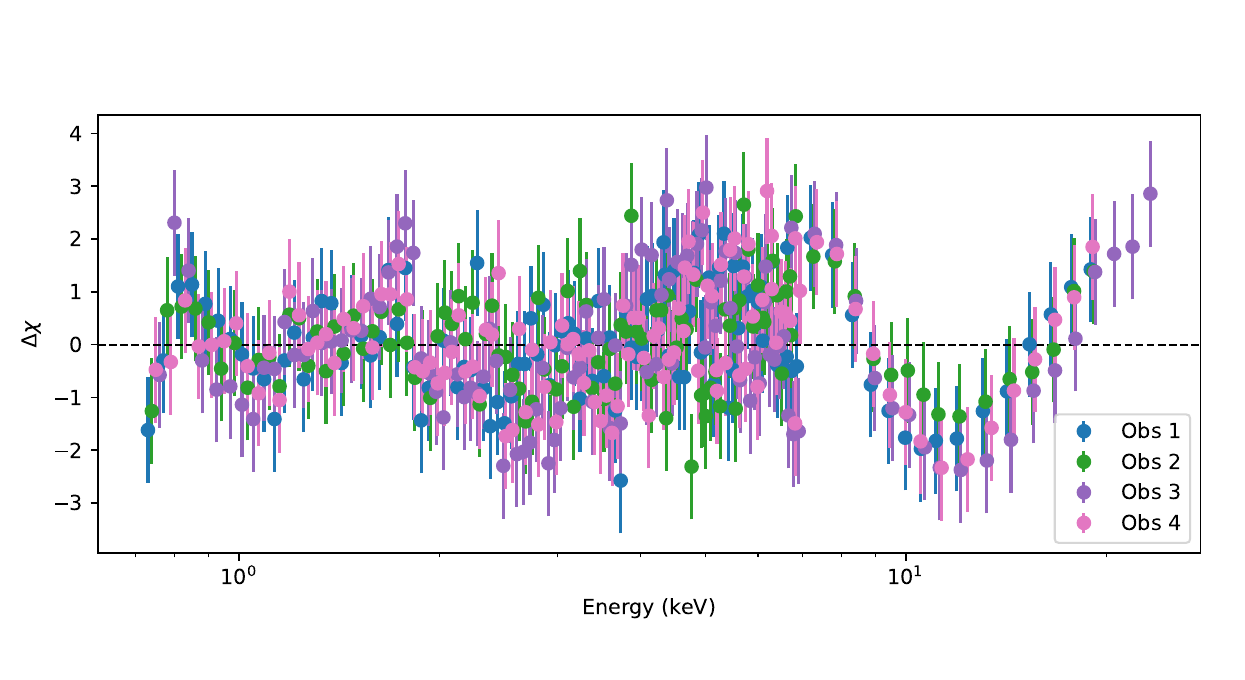}
	\end{tabular}
	\caption{The {$\Delta \chi$} of the spectral fitting using only
		{\tt Const * Tbabs * (Thcomp * bbodyrad)}. Excess around 6-10 keV
		can be seen out from this figure which needs to be taken care of. }
	\label{Fig01}
\end{figure}

The initial approach to model the spectra involves representing it as absorbed Comptonized emission, which arises from the inverse Comptonization of soft photons originating from the neutron star boundary layer. In terms of available XSPEC \citep{klitzing:1996ASPC..101...17A} routines, this model can be represented as {\tt Const * Tbabs * (Thcomp * bbodyrad)}. The constant factor accounts for the relative calibration between SXT and LAXPC. The combination of {\tt Thcomp} \citep{2020MNRAS.492.5234Z}, convoluted with {\tt bbodyrad}\footnote{\url{https://heasarc.gsfc.nasa.gov/xanadu/xspec/manual/node137.html}}, models thermal Comptonization, with soft blackbody radiation as the input. The {\tt Tbabs} \citep{klitzing:2000ApJ...542..914W} component accounts for intergalactic absorption. {The atomic abundance \citep{anders1989abundances} and the cross-sections \citep{1996ApJ...465..487V} have been set to the default XSPEC configuration.} Since {\tt Thcomp} is a convolution model, we had to extend the energy range for model computation to 0.1$-$1000 keV, using 500 logarithmic bins, to fit the spectra within the 0.7$-$25 keV range. This basic model combination yields a high reduced $\chi^2$ of approximately 1.35$-$1.45 across all observations, due to unresolved excess residuals in the 5$-$7 keV range (See Figure \ref{Fig01}). To address these residuals, we add a Gaussian component with its energy fixed at 6.5 keV. This adjustment leads to a significant improvement in the reduced $\chi^2$ value ($\sim$ 0.7$-$0.85 across all the observations) but the Gaussian width comes out to be very large $\sim$ 1.66$-$1.73 KeV. Such broadening can not be explained in terms relativistic effects considering an inferred blackbody (or disk) radius of $\sim$ 70 km (or $\sim$ 100 km) from a large normalisation of $\sim$ 9000, assuming a distance of 7.1 kpc. Furthermore, the inferred black body radius of 70 km is unreasonably large compared to the conventional radius of about 10 km, particularly within the context of the boundary layer emission.

\begin{figure}
	\centering
	\begin{tabular}{c}
		\includegraphics[scale=0.3]{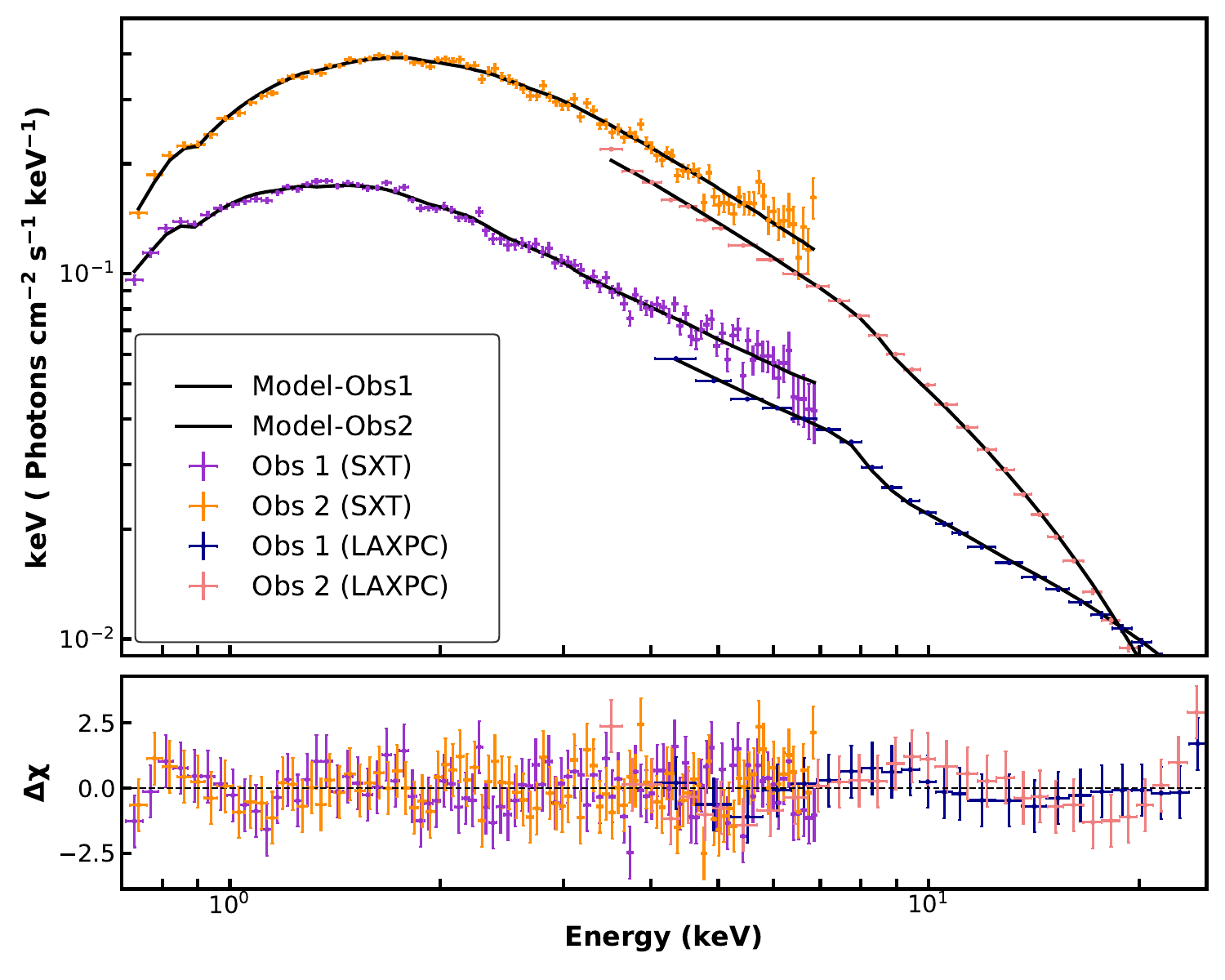}
	\end{tabular}
	\caption{{The spectra of Observation 1 (Hard) and Observation 2 (Soft). The blue and violet colour represents the SXT and LAXPC for Obs 1. Similarly, the orange and lightcoral shade represents the SXT and LAXPC for Obs 2. }}
	\label{Fig02}
\end{figure}

\begin{table*}
	\centering
	\renewcommand{\arraystretch}{1.7} 
	\setlength{\tabcolsep}{10pt}      
	\begin{tabular}{cccccc} 
		\hline
		{\bf Component} & {\bf Parameters} & {\bf Obs 1} & {\bf Obs 2} & {\bf Obs 3} & {\bf Obs 4} \\
		\hline
		\hline
		{\tt Tbabs} & $n_H \times 10^{22}$ (atoms/cm$^2$) & {{$0.19 \pm 0.02$}} & {{$0.25\pm 0.02$}} & {{$0.27\pm 0.02$}} & {{$0.27 \pm 0.03$}}\\
		\hline
		{\tt Thcomp} & Photon Index ($\Gamma$) & {{$2.35^{+0.02}_{-0.02}$}} & {{$2.46^{+0.09}_{-0.09}$}} & {{$2.32^{+0.09}_{-0.10}$}} & {{$2.32^{+0.05}_{-0.05}$}} \\
		& Electron Temp ($kT_e$) (keV) & {{$18.2^{+7.7}_{-4.1}$}} & {{$4.0^{+0.41}_{-0.33}$}} & {{$3.15^{+0.23}_{-0.19}$}} & {{$6.1^{+0.87}_{-0.79}$}}\\
		\hline 
		{\tt bbodyrad} & Black body Temp ($kT_{bb}$) (keV) & {{$0.72^{+0.01}_{-0.01}$}} & {{$0.90^{+0.02}_{-0.02}$}} & {{$0.75^{+0.01}_{-0.01}$}} & {{$0.75^{+0.01}_{-0.01}$}}\\
		\hline
		{\tt Disk} & Inner Disk Temp ($kT_{in}$) (keV) & {{$0.50^{+0.03}_{-0.03}$}} & {{$0.72^{+0.03}_{-0.03}$}} & {{$0.63^{+0.03}_{-0.03}$}} & {{$0.57^{+0.02}_{-0.02}$}} \\
        \hline
		{\tt relconv } & {\tt $R_{in}$} in units of ($R_g = \frac{GM}{c^2}$) & {{$8.3^{+1.22}_{-1.06}$}} & {{$6.3^{+0.62}_{-0.57}$}} & {{$5.4^{+0.8}_{-0.6}$}} & {{$8.1^{+2.0}_{-1.4}$}}\\
		\hline
        {\tt xilconv } & {Rel$_{refl}$} & {{$2.20^{+0.65}_{-0.89}$}} & {{$0.65^{+0.58}_{-0.48}$}} & {{$1.31^{+0.66}_{-0.60}$}}& {{$1.67^{+0.71}_{-0.44}$}}\\
        & {$\log \xi$} & {{$1.72^{+0.19}_{-0.48}$}} & {{$2.30^{+0.35}_{-0.33}$}}& {{$2.03^{+1.61}_{-0.62}$}} & {{$2.99^{+0.19}_{-0.17}$}}\\
        \hline
		& $\chi^2$/dof & {{77.4/107}} & {{102.3/116}} & {{89.5/106}} & {{77.5/111}}\\
		\hline
	\end{tabular}
	\caption{Spectral best fitting parameters for four observations with the model {\tt Const * Tbabs * (relconv * xilconv * Thcomp * bbodyrad + diskbb )}. All error values are calculated at the 90 \% confidence level. $\dagger$ represents the particular value which remains un$-$constrained during the fitting. During all the fittings, we have taken the limb darkening factor to be 0, meaning isotropic brightening.}
	\label{Table2}
\end{table*}

We adopted a similar approach to that discussed in \citet{Chattopadhyay_2024} to model the spectra of 4U 1702$-$429 (Ara X$-$1) in the soft states using a three$-$component model. We have modelled the spectra considering the combination: {{\tt Const * Tbabs * (relconv * xilconv * Thcomp * bbodyrad + diskbb )}. To speed up the model, calculation of the output spectrum, we have used {\tt xset} to define max energy to be 500 keV in the present scenario.}  Thus, in this scenario the reflection component, which is taken care of by {{\tt xilconv} routine which takes the incident radiation coming from thermal Comptonizing corona while the Comptonizing medium itself takes input flux from the soft black body component. While {\tt xilconv} \citep{2011MNRAS.416..311K} takes care of both the ionised disk along with the Compton reflection, {\texttt{relconv} \citep{2010MNRAS.409.1534D} accounts for relativistic blurring and the distortion of spectral features caused by relativistic effects, including Doppler shifts, gravitational redshift, and boosting.}}  It is to be noted that eliminating the {\tt diskbb} \citep{1984PASJ...36..741M} component results in a larger black body normalization. Thus, we had to fix the black body normalisation to 198, assuming a typical radius of 10 km at a distance of 7.1 kpc and added a disk component to take care of the additional soft residuals that yet to be fixed. During the fitting, the best fitted value of the covering fraction reached to its maximum allowed value 1, hence we have fixed it at 1 during all the fittings.  Moreover, we have tied the normalisation of the disk component to the inner radius parameter of the {\tt xilconv} component and kept the inner radius free. The best fitted parameters are tabulated in the Table \ref{Table2}. We note that the reduced $\chi^2$ quoted in Table \ref{Table2} is lower than unity, which is due to the conservative systematic error of 3\% being used. If instead the systematic error is set at 2\%, the reduced $\chi^2$ is near unity. It is important to note that during the spectral fitting, we fixed the spin parameter ($\tilde{a}$) to 0.27 based on earlier reports \citep{10.1093/mnras/stx671}. The outer disk radius and the break radius have been set to 400 $r_g$ and 399 $r_g$, respectively. Fe abundance and inclination angle have been kept fixed at 1 and 74 $\deg$ based on the earlier report \citep{10.1093/mnras/stx671}. The emissivity index are set to the default value 3. Considering the best fitting parameters of the {\tt R$_{in}$} we get an estimate of inner disk radius of $\sim$ 8$R_g$ in the hard state and $\sim $ 6 $R_g$  in the relatively softer state. The Figure \ref{Fig02} displays the spectra for Obs 1 and Obs 2, representing the extreme hard and extreme soft state among the four observations.

\section{Timing Analysis} \label{sec5}

The Large Area X$-$ray Proportional Counter (LAXPC) onboard {\em AstroSat}, with its temporal resolution of 10 $\mu s$ and an effective area of 6000 cm$^2$ at 15 keV, covering the 3$-$80 keV energy range, offers an excellent opportunity to study the source temporal behaviour. Using all the PCUs, we have generated the PDS for the 3$-$20 keV range, employing the standard routine {\tt laxpc\_find\_freqlag}, with a Nyquist frequency of 2000 Hz corresponding to a light curve with a bin time of 0.25 milliseconds for all observations. Light curves are split into small segments of  8.192 s, and the PDS from those small segments are averaged to get the final PDS. Thus, in each scenario, the minimum frequency for the PDS comes out to be 0.122 Hz. Further, these un$-$binned PDS have been re$-$binned using the routine {\tt laxpc\_rebin\_power}  with a signal-to-noise ratio of 5. The routine produces the PDS that is dead time (42 $\mu $s)  corrected and Poisson noise subtracted. However, we have detected the presence of faint residuals at high frequencies present in the PDS, which has been taken care of using a constant factor in the modeling.

\begin{table*}
	\centering
\renewcommand{\arraystretch}{1.5} 
\setlength{\tabcolsep}{16pt}      
\begin{tabular}{cccccc} 
    \hline
    {\bf Component} & {\bf Parameter} & {\bf Obs 1} &  {\bf Obs 2} & {\bf Obs 3} & {\bf Obs 4} \\
    \hline
    \hline
    {\tt Powerlaw}     & Index & {0}${^\dagger}$ & {0}${^\dagger}$ & {0}${^\dagger}$ & {0}${^\dagger}$ \\
    & N$_p\times 10^{-5}$ & {10.4 $\pm$ 1.7} & {3.27 $\pm$ 0.10} & {5.10 $\pm$ 0.17} & {7.2 $\pm$ 1.2} \\
    \hline              
    {\tt Lorentzian 0} & $\nu_0$ & {0}${^\dagger}$ & {0}${^\dagger}$ & {0}${^\dagger}$ & {0}${^\dagger}$ \\
    & $\sigma_0$ & {38 $\pm$ 3} & {4.7 $\pm$ 4.5} & {0.11 $\pm$ 0.05} & {$<$ 0.73} \\
    & RMS (\%) & {17 $\pm$ 0.4} & {2.0 $\pm$ 0.4} & {2.0 $\pm$ 0.2} & {1.1 $\pm$ 0.3} \\  
    \hline
    {\tt Lorentzian 1} & $\nu_1$ & {197}${^\dagger}$ & {151 $\pm$ 16}  & {150}${^\dagger}$ & {116}${^\dagger}$ \\
    & $\sigma_1$ & {2120 $\pm$ 470}  & {86 $\pm$ 47} & {$>$ 810} & {1700 $\pm$ 600} \\
    & RMS (\%) & {50 $\pm$ 9} & {5.0 $\pm$ 0.8} & {15 $\pm$ 1} & {31 $\pm$ 6} \\      
    \hline
    {\tt Lorentzian 2} & $\nu_2$ & -- & -- & {0.6 $\pm$ 0.4} & {13 $\pm$ 2} \\
    & $\sigma_2$ & -- & -- & {1.2 $\pm$ 0.8} & {55 $\pm$ 9} \\
    & RMS (\%) & -- & -- & {1.8 $\pm$ 0.4} & {37 $\pm$ 0.2} \\  
    \hline
    {\tt Lorentzian 3} & $\nu_4$ & -- & {36 $\pm$ 8} & {29 $\pm$ 4} & -- \\
    & $\sigma_4$ & -- & {40 $\pm$ 20} & {39 $\pm$ 12}  & -- \\
    & RMS (\%) & -- & {5.6 $\pm$ 1.9} & {5.2 $\pm$ 0.5} & -- \\  
    \hline  
    {\tt Lorentzian 4} & $\nu_5$ & {644 $\pm$ 11} & {716 $\pm$ 3} & {619 $\pm$ 19} & {850 $\pm$ 10} \\
    & $\sigma_5$ & {121 $\pm$ 37} & {46 $\pm$ 8} & {43 $\pm$ 41} & {183 $\pm$ 58} \\
    & RMS (\%) & {12.8 $\pm$ 1.3} & {7.9 $\pm$ 0.4} & {3.7 $\pm$ 1.3} & {13.4 $\pm$ 0.2} \\ 
    \hline 
    {\tt Lorentzian 5} & $\nu_6$ & -- & {972 $\pm$ 15} & {882 $\pm$ 8} & --  \\
    & $\sigma_6$ & -- & {127 $\pm$ 63} & {132 $\pm$ 41} & -- \\
    & RMS (\%) & -- & {6.3 $\pm$ 1.0} & {9.9 $\pm$ 0.9} & -- \\ 
    \hline   
    & $\chi^2$ /dof & {146/139} & {145/107} & {122/121} & {160/126} \\ 

		\hline                       	                          		 
	\end{tabular}
	\caption{Details of the best fit parameters we obtained from the PDS fitting using multi multi-Lorentzian combination for Obs 1$-$Obs 4. All the errors reported here have been calculated at 90 \% confidence level. $\nu$ and $\sigma$ represent the centroid frequency, FWHM, respectively. The number in the suffix of each of the components stands for representing that particular Lorentzian component. $\nu$ and $\sigma$ has the unit of Hz. The rms has been computed from the square root of the normalisation of the Lorentzian, which is reported in \%, and $N_p$ represents the power law normalisation.}
	\label{Table3}	
\end{table*}

\begin{figure*}
	\centering
	\begin{tabular}{cc}
		\includegraphics[scale=0.3]{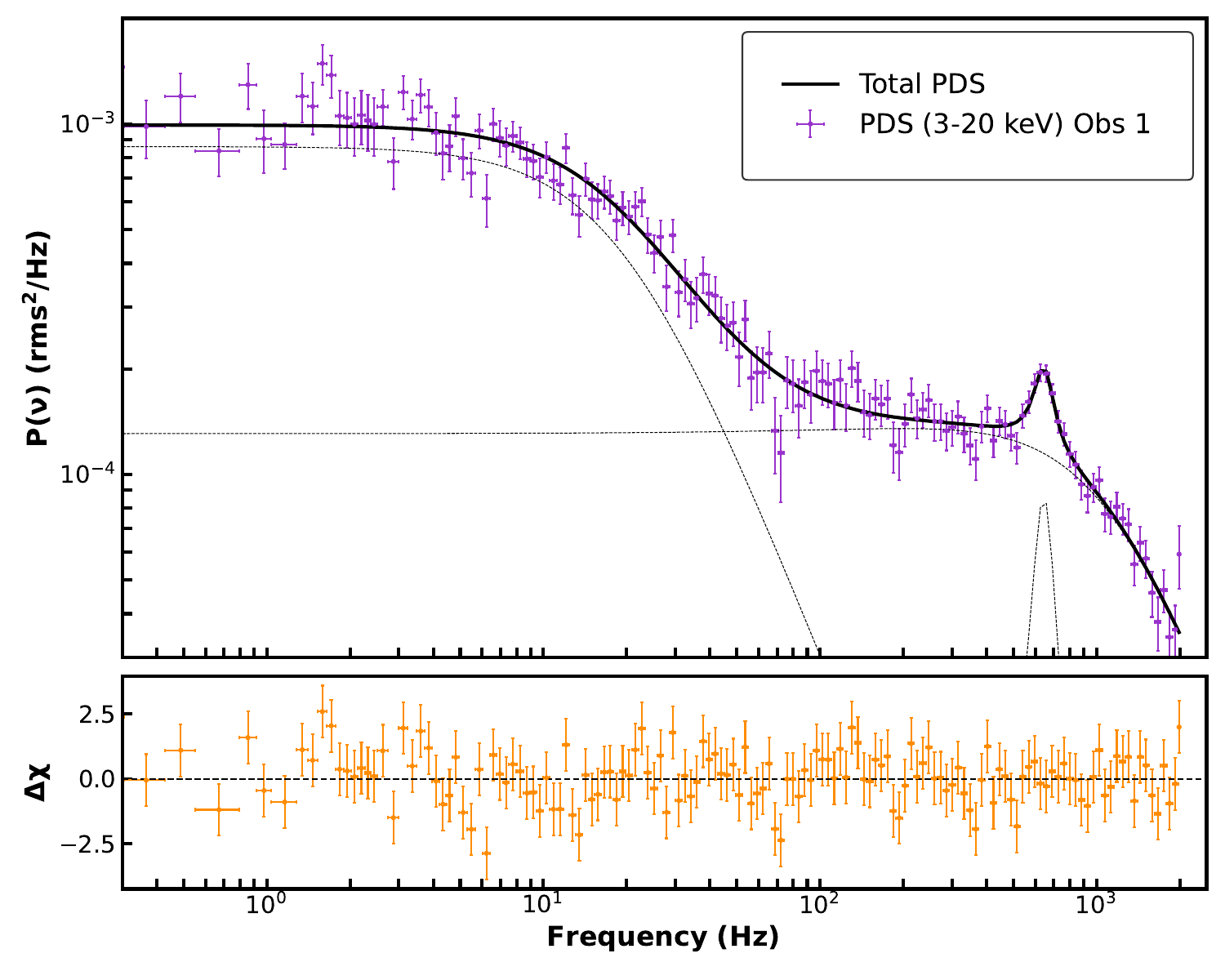} & \includegraphics[scale=0.3]{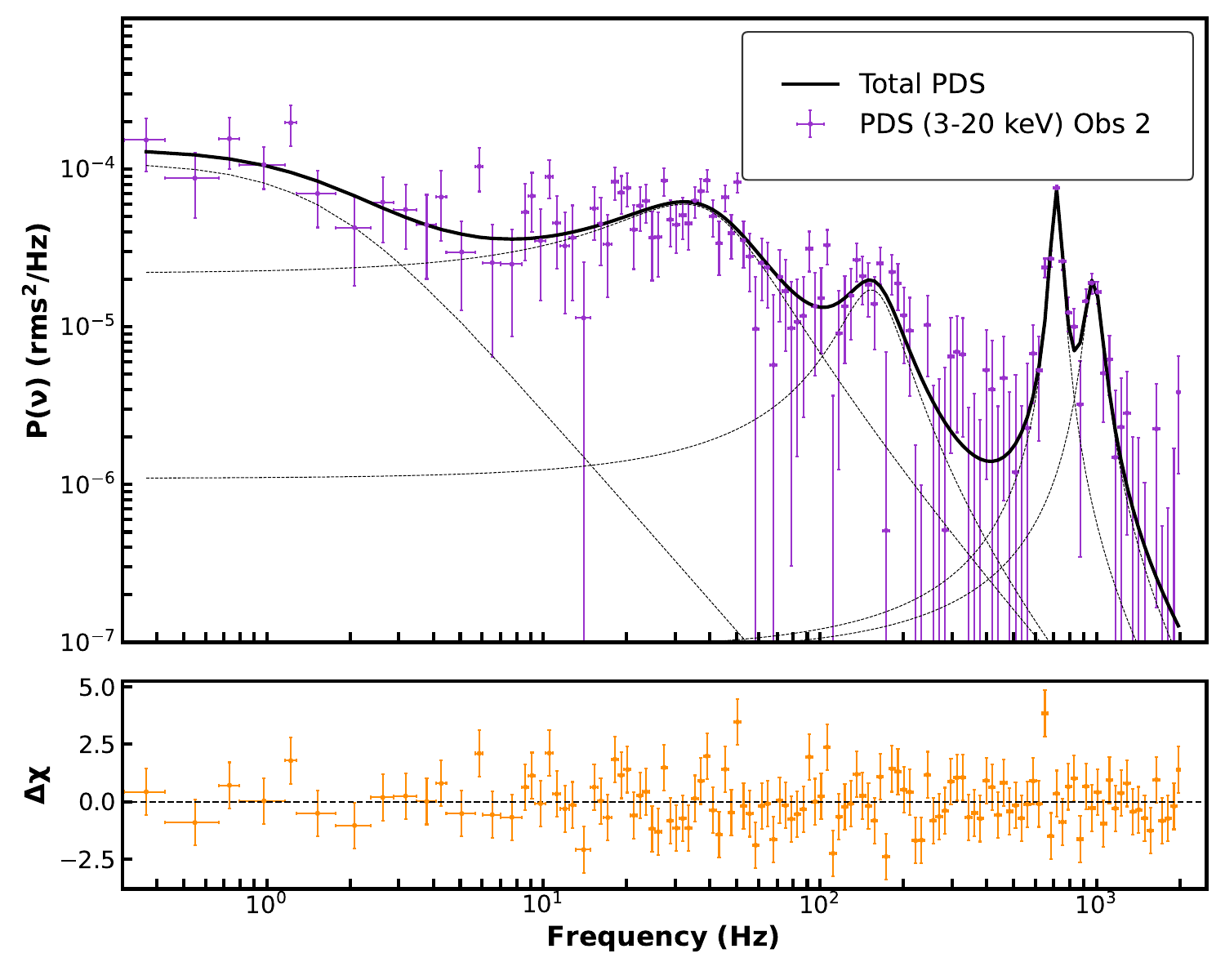}\\
		\includegraphics[scale=0.3]{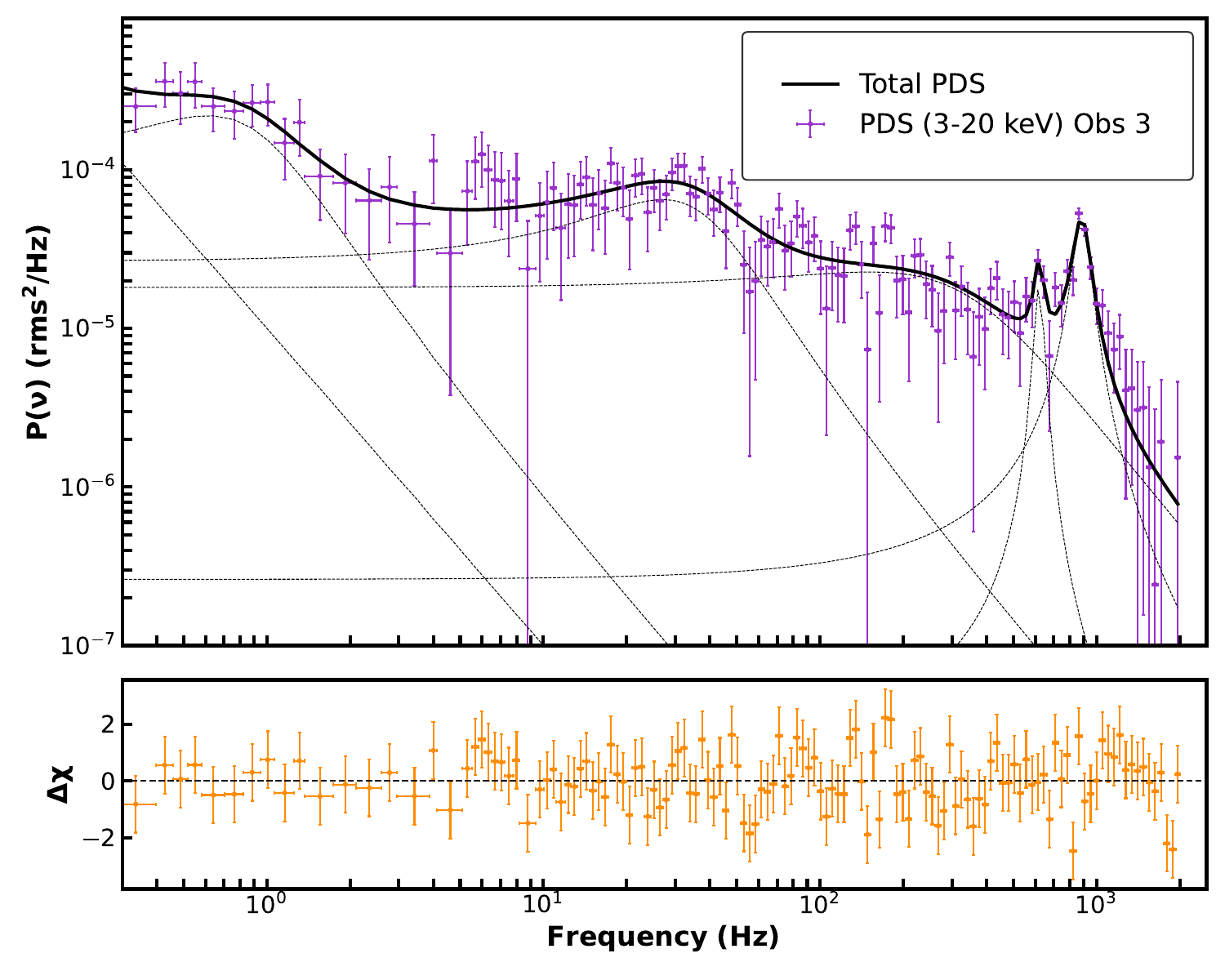} & \includegraphics[scale=0.3]{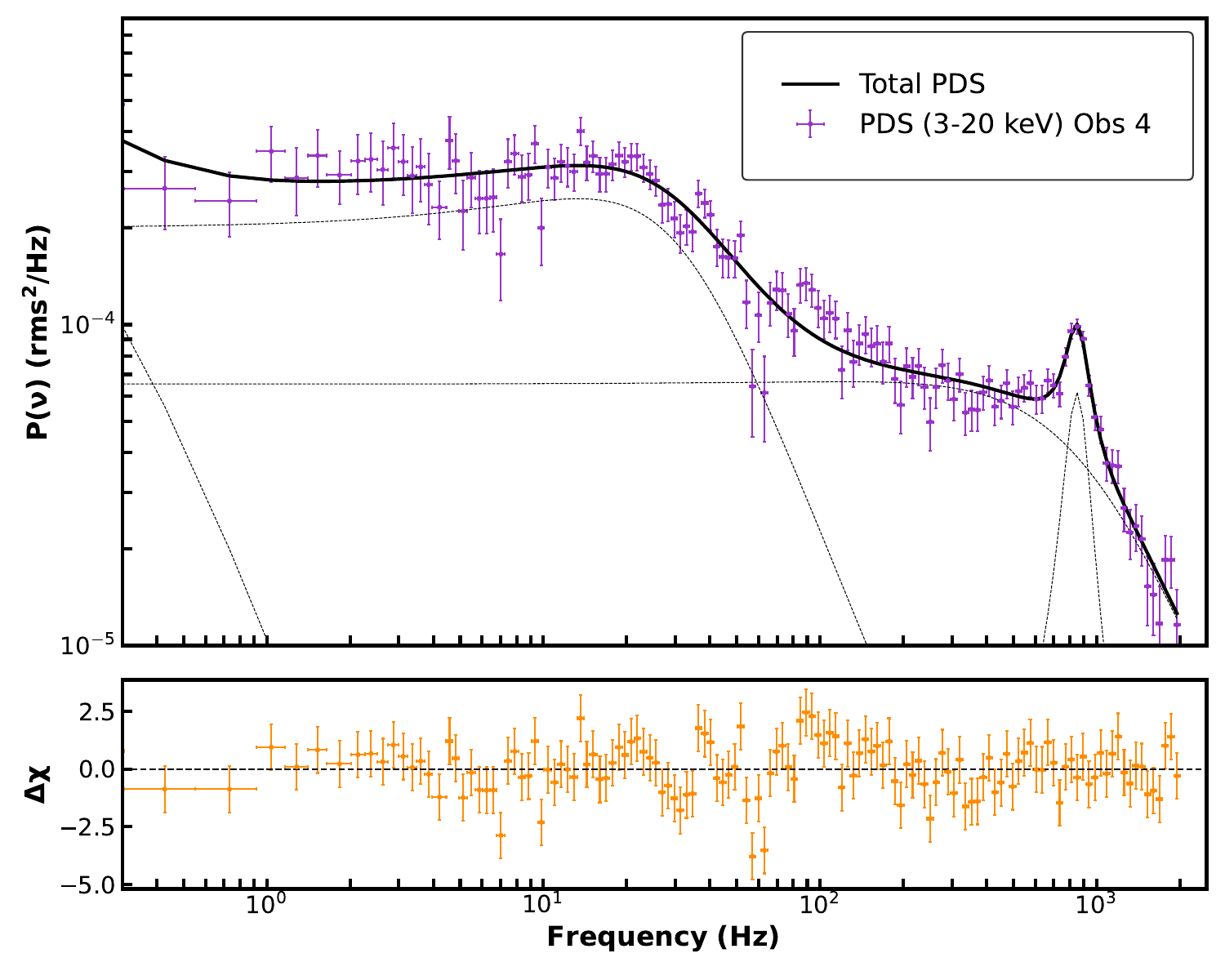}\\
	\end{tabular}
	\caption{ PDS of 4U 1636$-$536 from Obs$-$1 to Obs$-$4 with in 3$-$20 keV range employing all the PCUs. Obs 2 and Obs 3 show a clear presence of twin QPOs where as Obs 1 has a signature of a lower kHz QPO, and Obs 4 has a signature of an upper kHz QPO.}
	\label{Fig03}
\end{figure*}

We modelled the PDS using multiple Lorentzian components. The best-fitted parameters have been tabulated in the Table \ref{Table3}. The best-fitted parameters indicate to the detection of two single kHz QPOs in Obs 1 and Obs 4, while twin peaks have been detected in Obs 2 and Obs 3, which are relatively softer. The presence of the $\sim$ 28 Hz to 34 Hz broad hump becomes prominent with the presence of the { twin kHz QPO} peaks. Additionally, a noticeable broad peak around $\sim$ 200\,Hz is consistently observed in the PDS, becoming more prominent in the extreme soft state with a {$Q$ of $sim$ 1.75}. 
The best fitted PDSs has been presented in the Figure \ref{Fig03}. 

\subsection{Time Lag and rms calculation}

\begin{figure}
	\centering
	\includegraphics[scale=0.3]{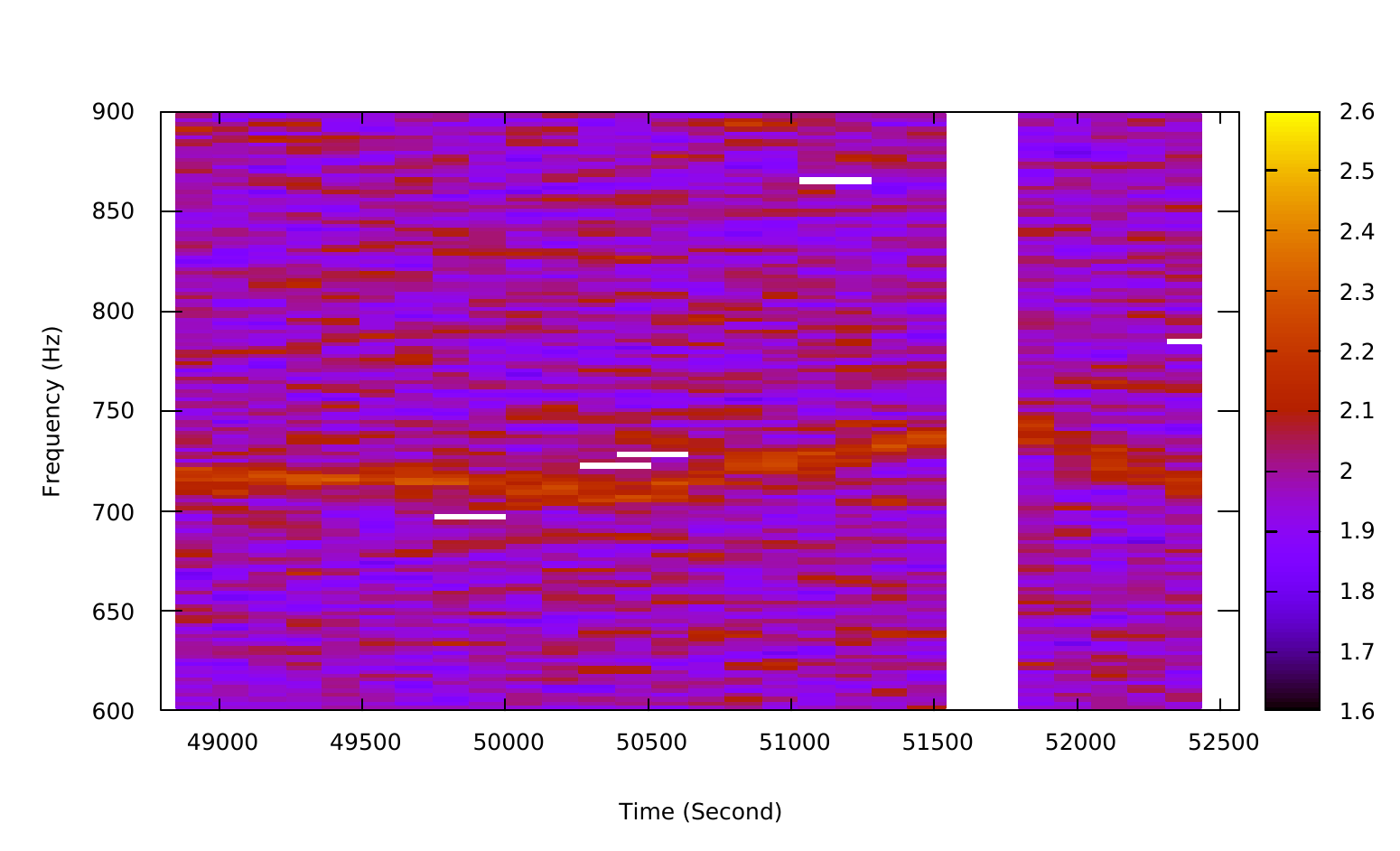} 
	\caption{The dynamic power density spectrum of 4U 1636$-$536 in Obs 2, captured between 49,000 and 52,500 seconds, is shown. The right-hand axis of the figure represents the Leahy-normalised power.}
	\label{Fig05}
\end{figure}

The routine {\tt laxpc\_find\_freqlag} enables us to compute the energy$-$dependent time lag and fractional rms {spectra} at the QPO frequency as well. The dynamic power density spectra using Nyquist Frequency of 2000 Hz, frequency resolution of 2 Hz and time segment of 128 second across various observations reveals that the QPO frequency fluctuates throughout the observation period. For instance, Figure \ref{Fig05} illustrates the frequency evolution of the kHz QPO over time in a specific segment of Obs 2. To achieve more accurate energy$-$dependent variability, we have calculated the lag and fractional rms spectra specifically in segments of the good time interval where the QPO frequency remains relatively stable. 

\begin{figure*}
	\centering
	\begin{tabular}{c c c}
		Obs 1 &\includegraphics[scale=0.3]{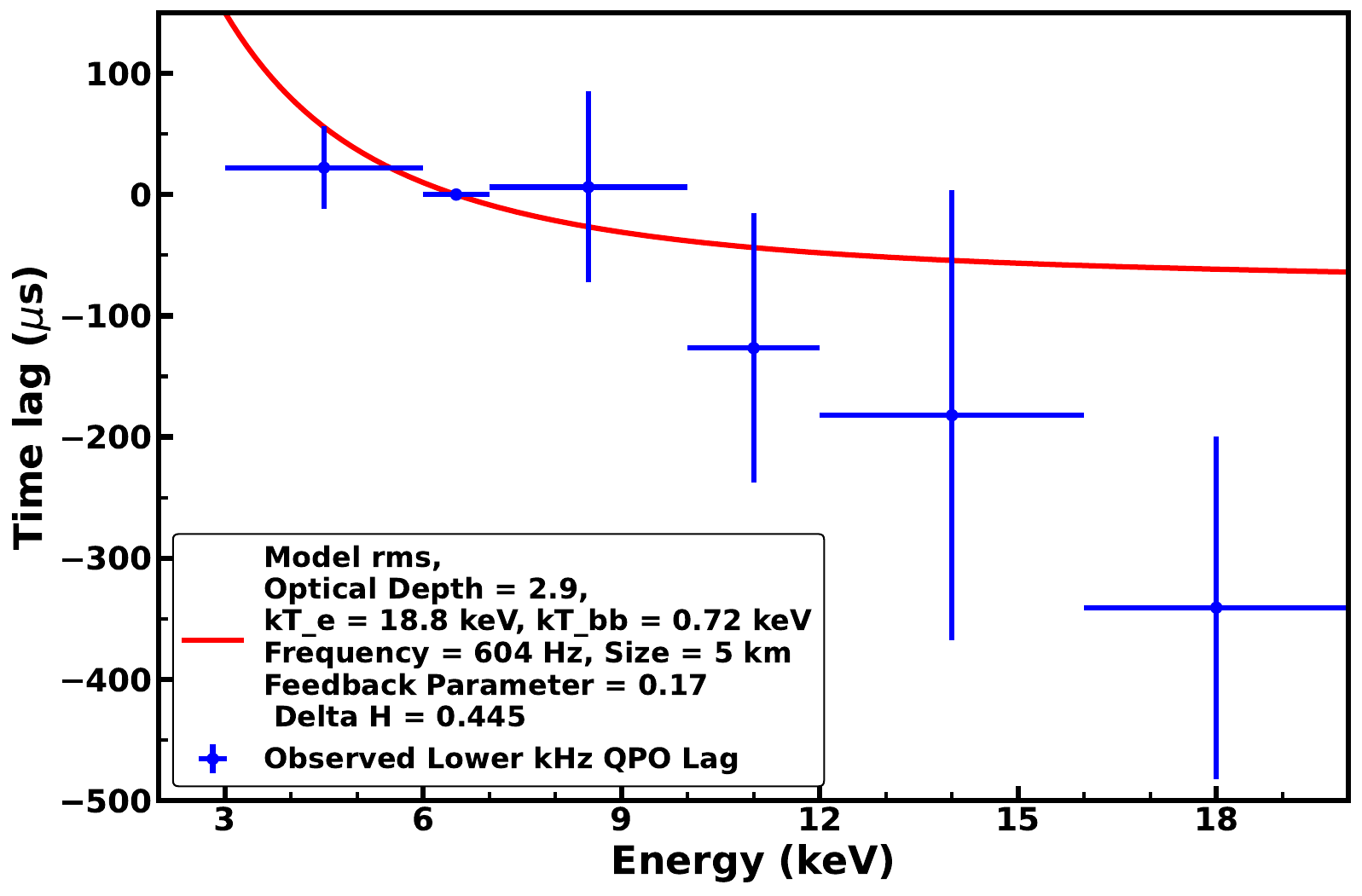} & \includegraphics[scale=0.3]{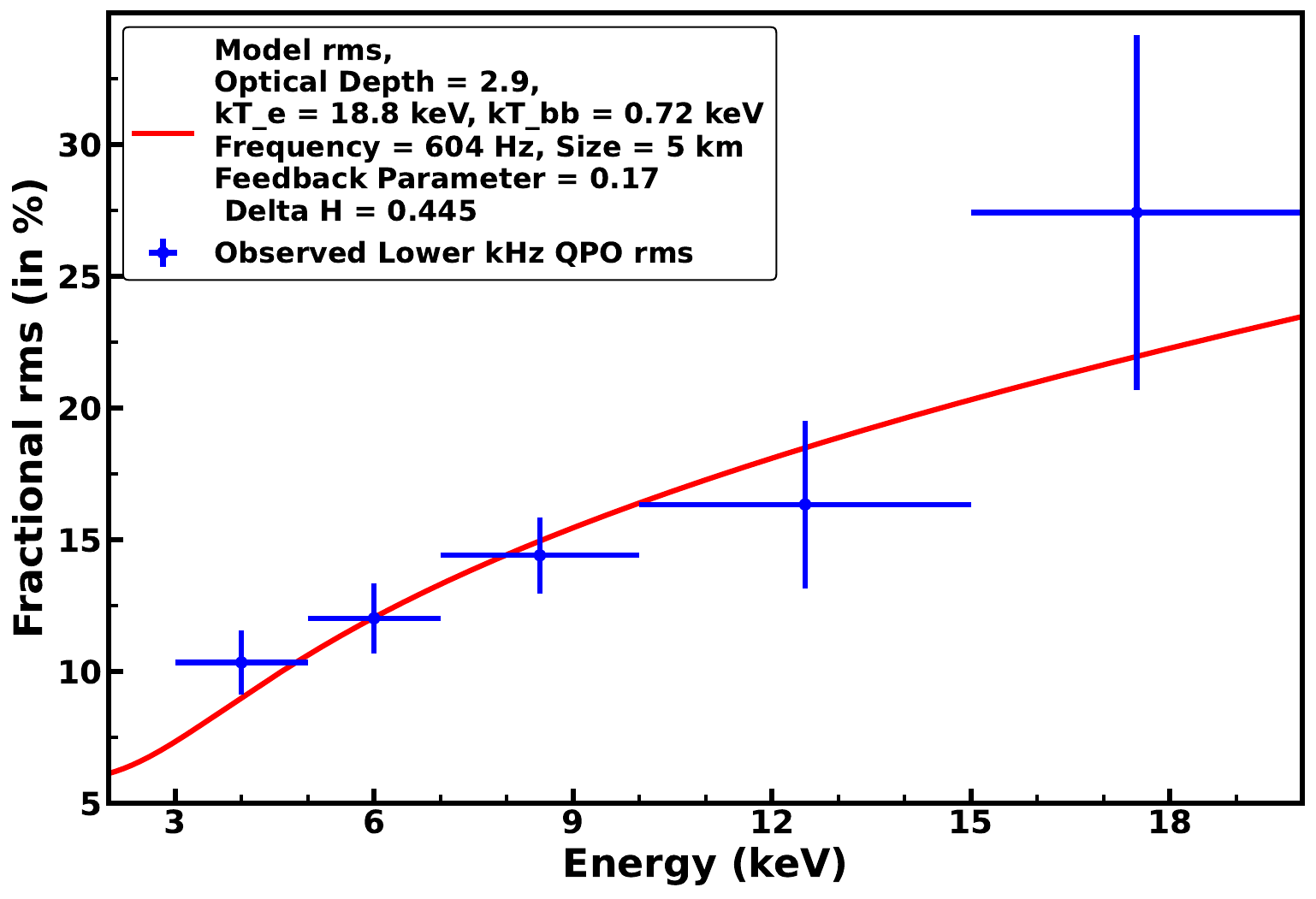}\\
		Obs 2 &\includegraphics[scale=0.3]{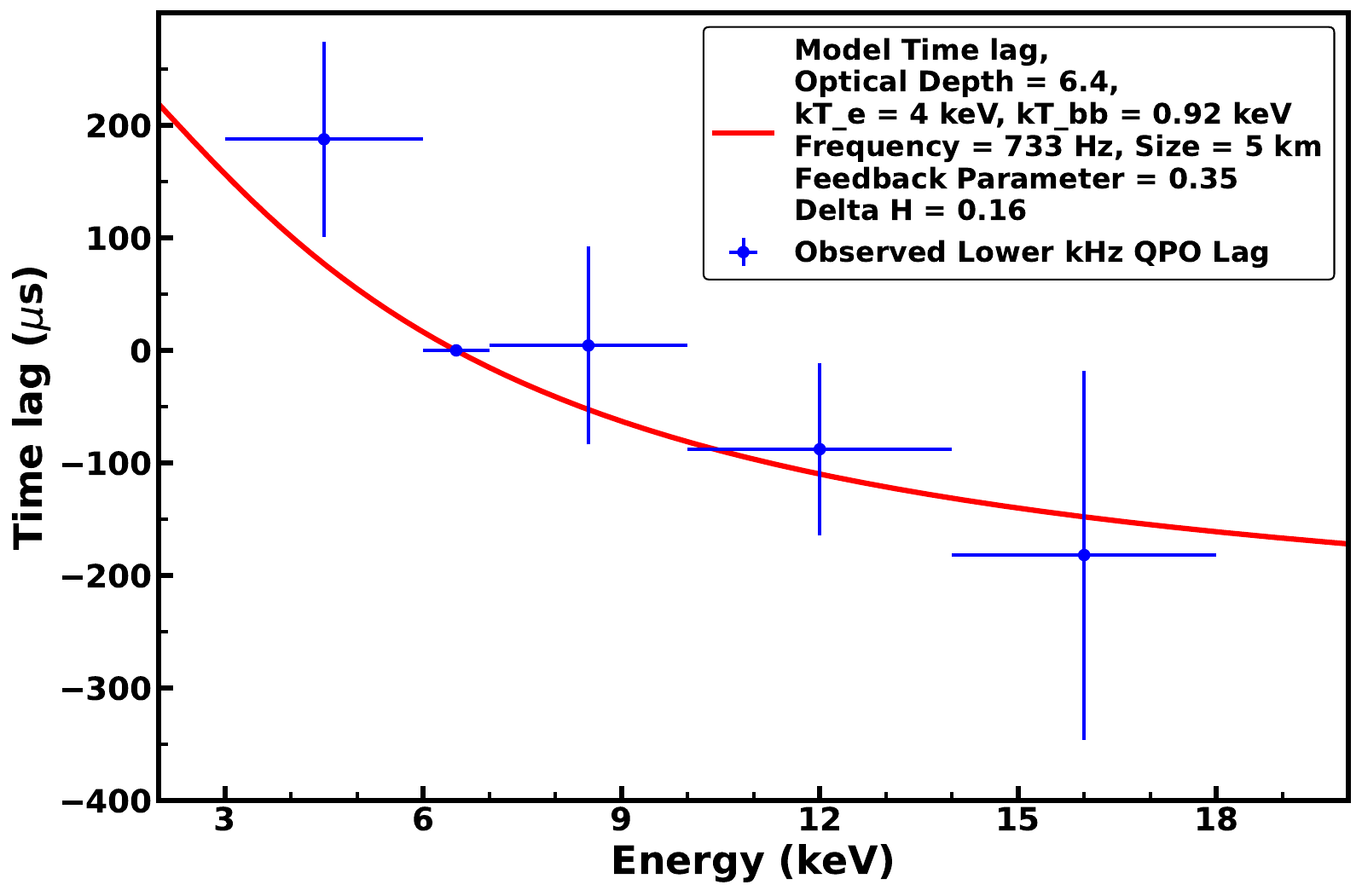} & \includegraphics[scale=0.3]{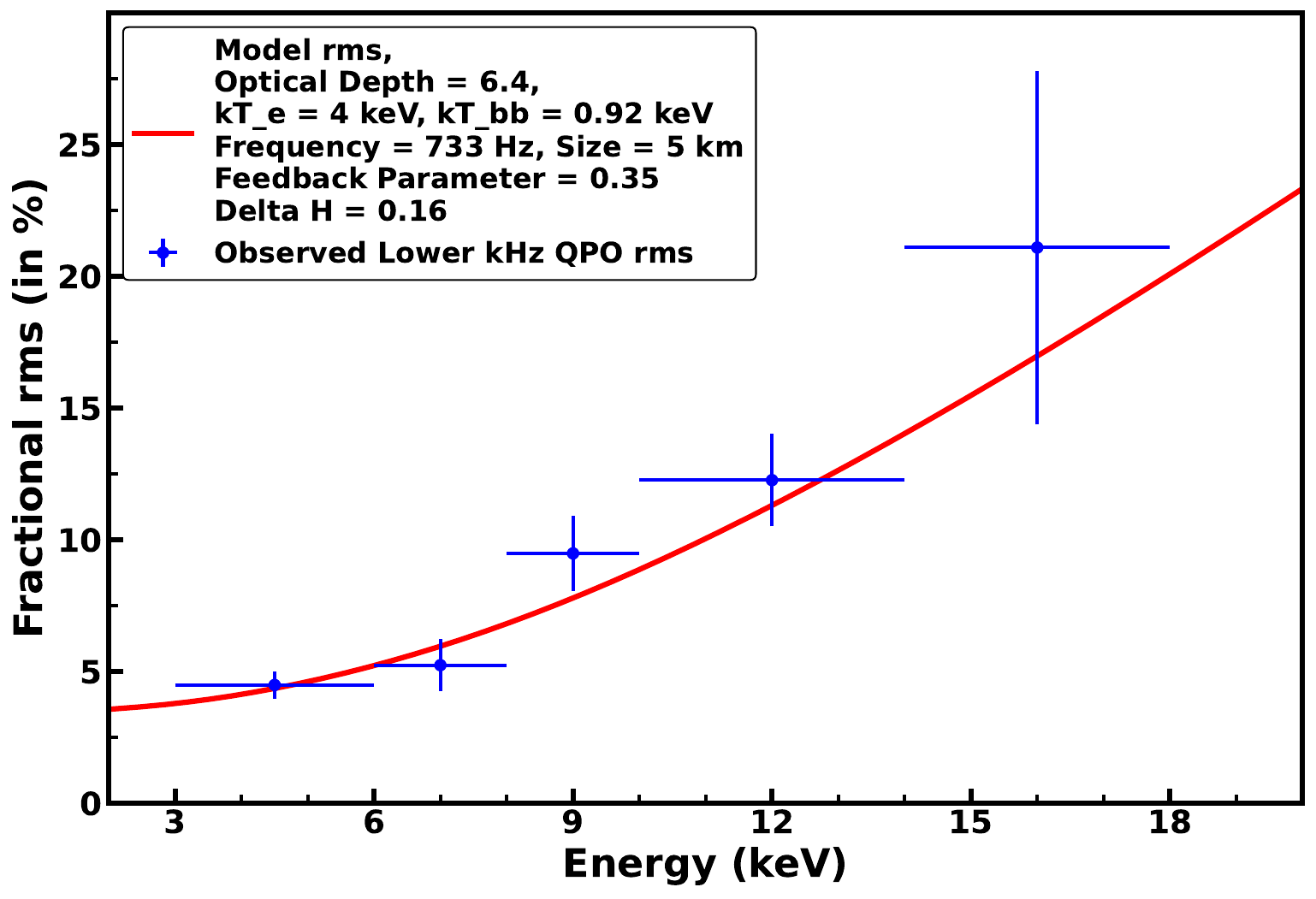}\\
		Obs 3 &\includegraphics[scale=0.3]{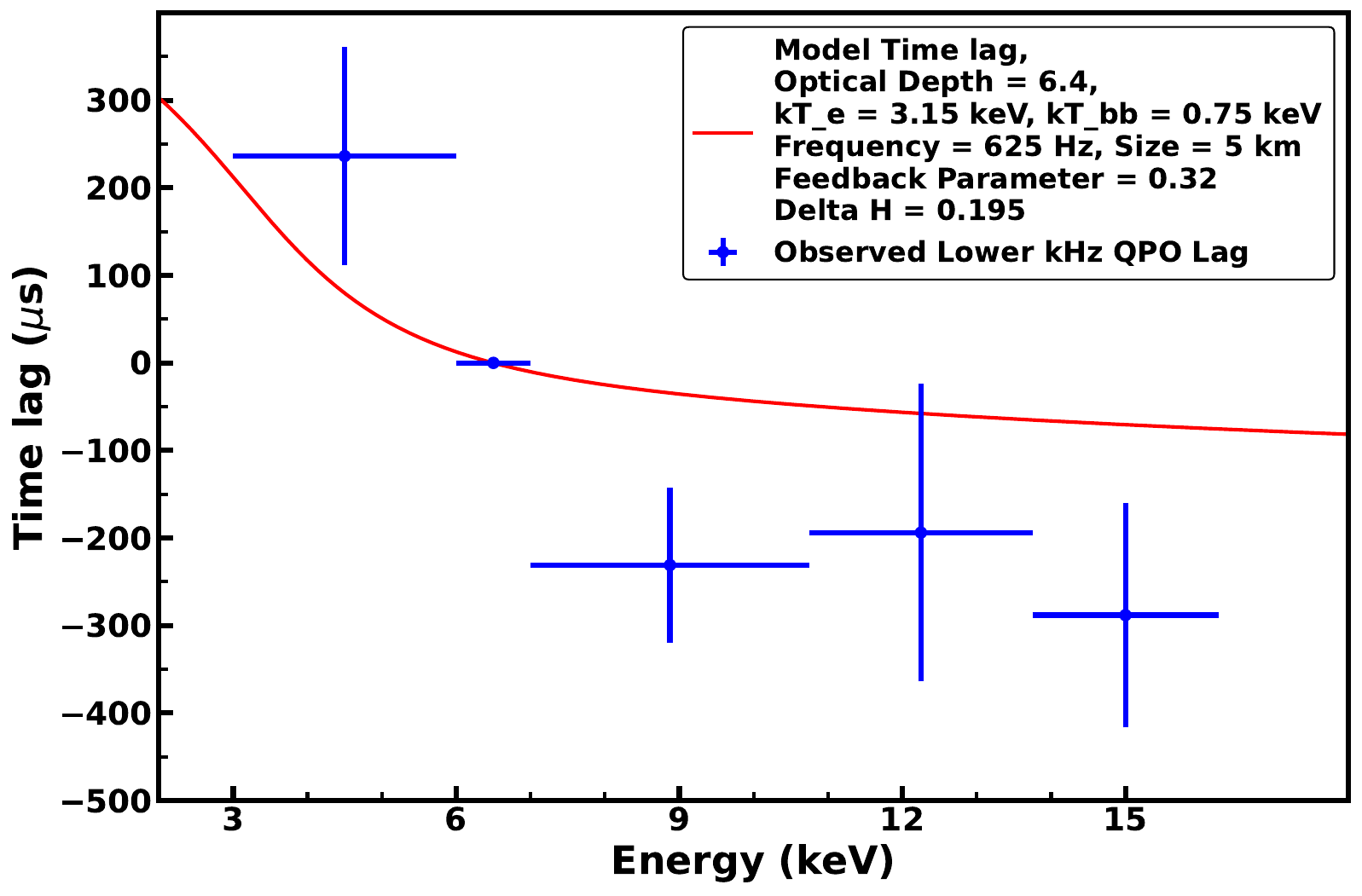} & \includegraphics[scale=0.3]{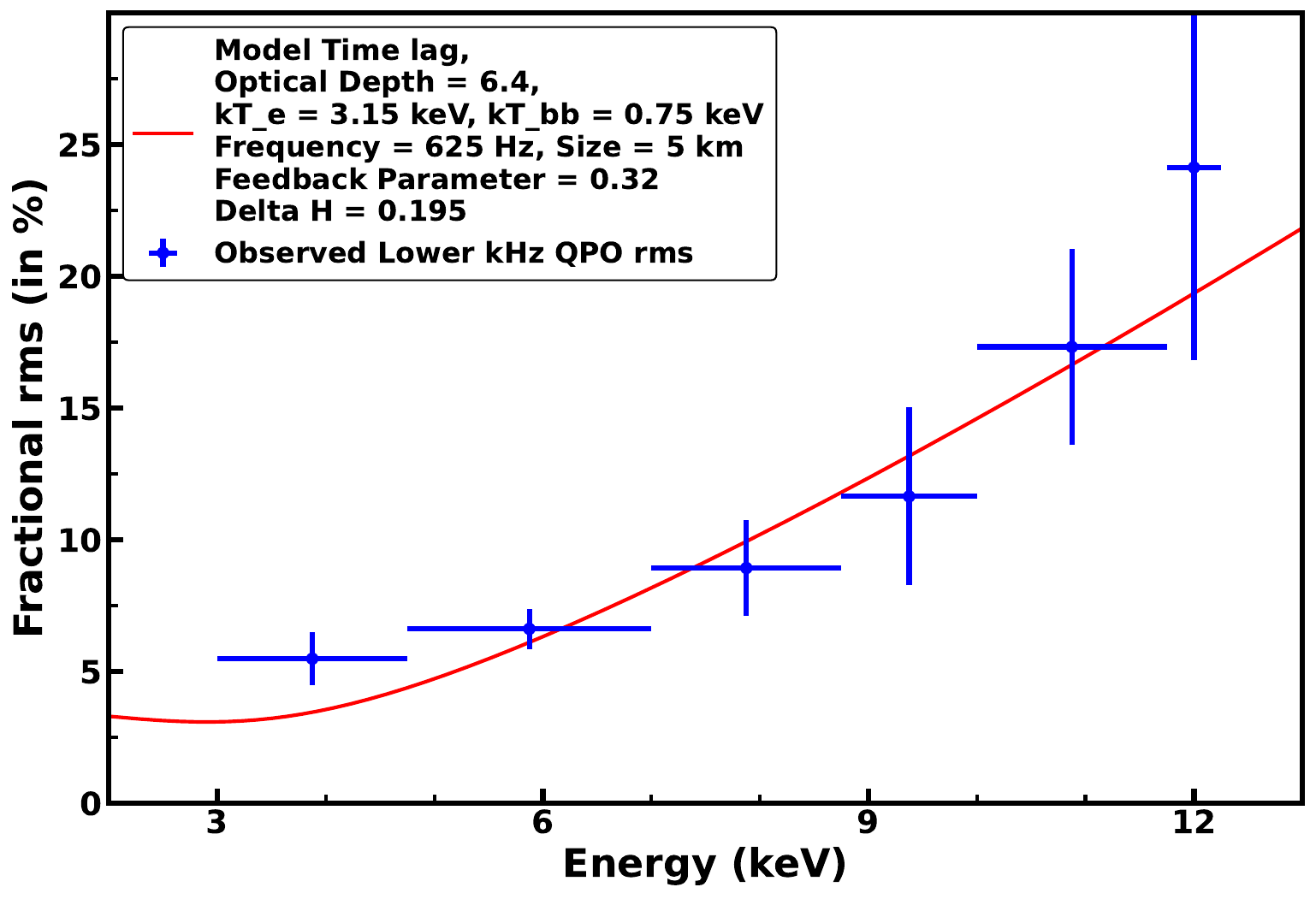}\\
		Obs 4 &\includegraphics[scale=0.3]{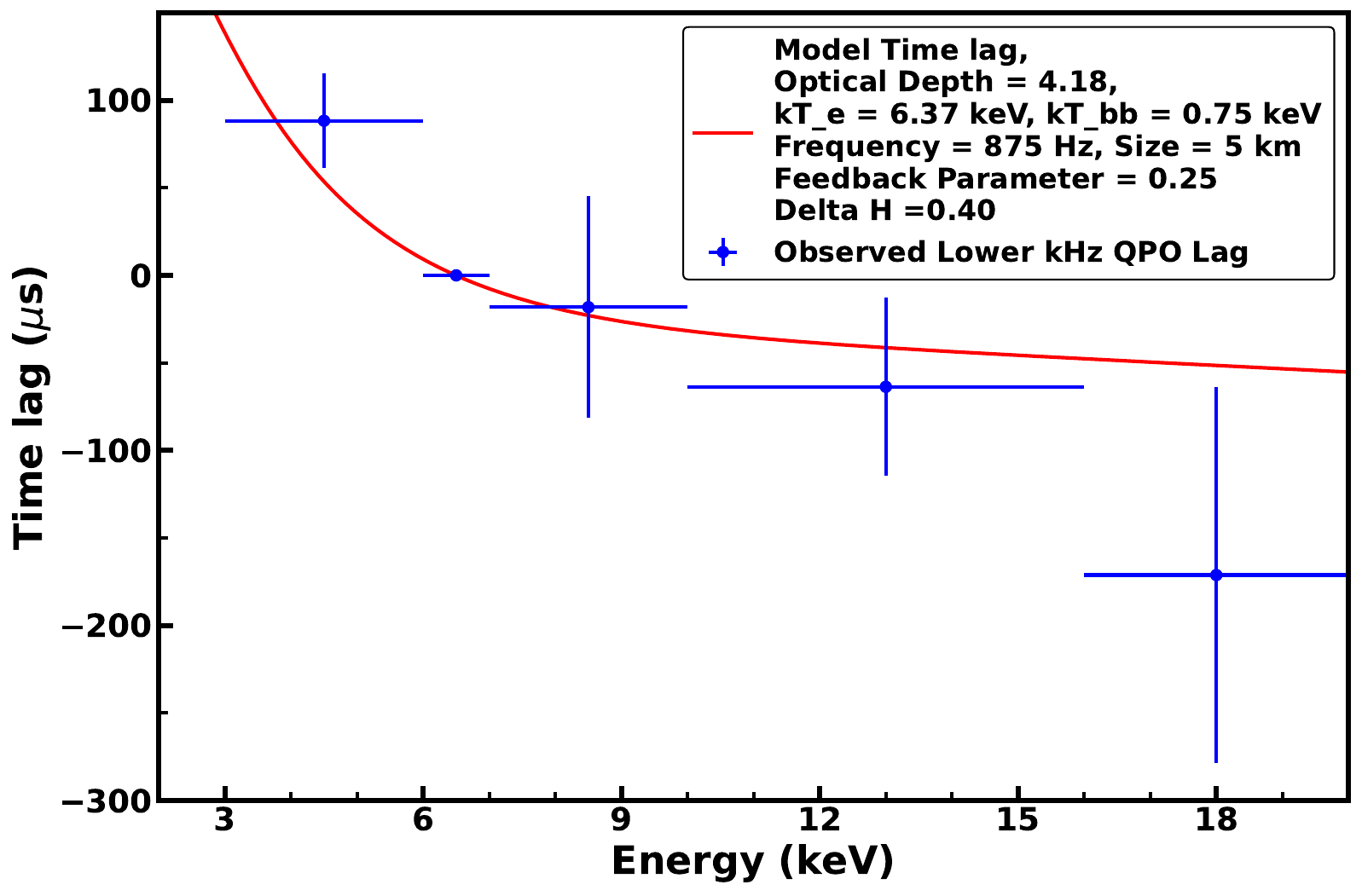} & \includegraphics[scale=0.3]{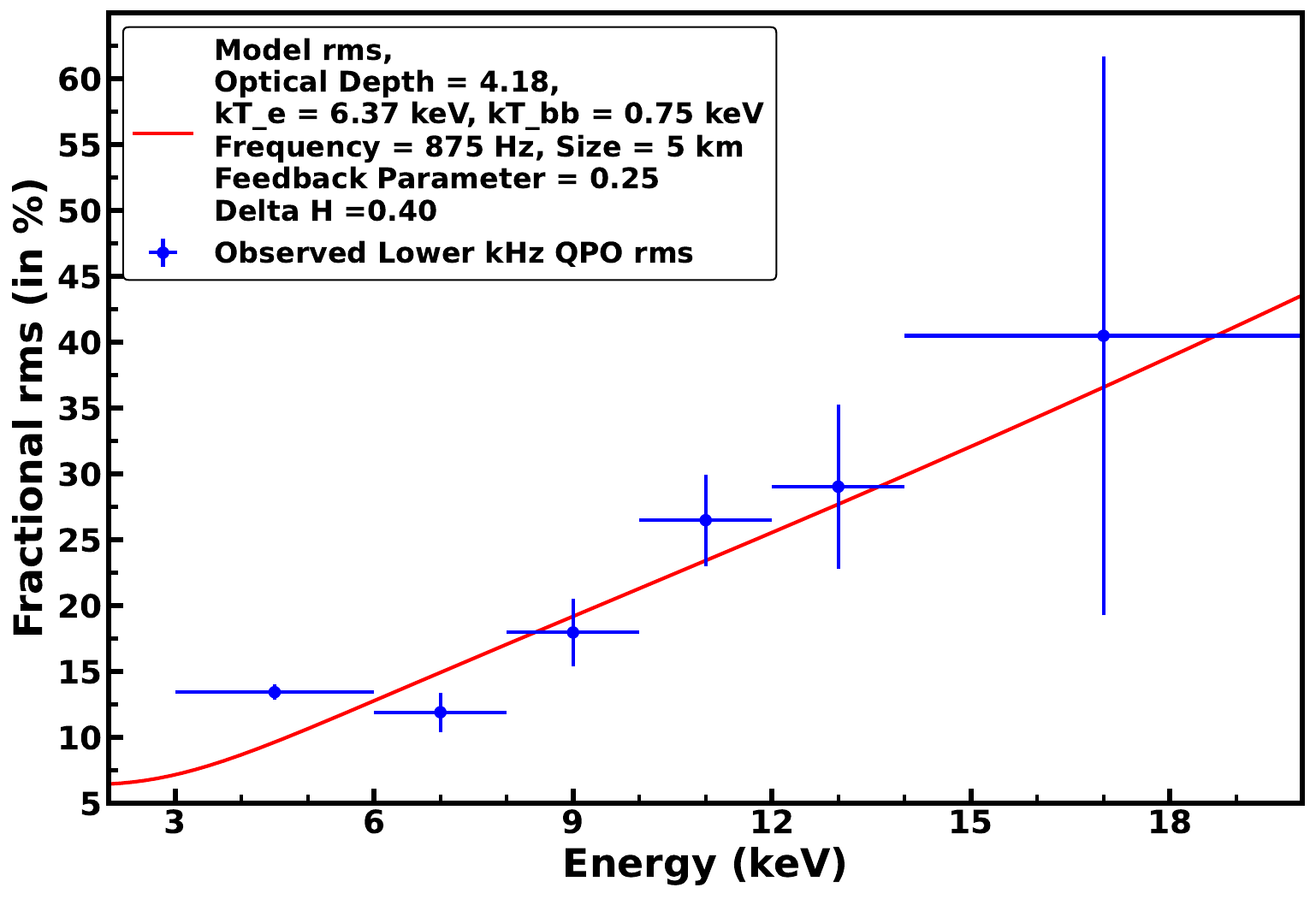}\\
	\end{tabular}
	\caption{ Energy dependent time$-$lag and fractional rms spectra in different observation at lower kHz QPO frequency for 4U 1636$-$536. The red solid line represents the theoretical model prediction of the rms and lag using Comptonizing model of lag calculation with the given spectral parameters. }
	\label{Fig06}		
\end{figure*}

We have calculated the energy$-$dependent fractional rms and time lag spectra using a frequency resolution equivalent to the QPO width, incorporating data from all PCUs. This allows the routine {\tt laxpc\_find\_freqlag} to compute lags and fractional rms within the frequency range from $\nu_0 $-$ \Delta \nu$ to $\nu_0 + \Delta \nu$. For all scenarios, the lags for lower kHz QPOs have been calculated with respect to the reference energy band of 6$-$7 keV, while for {LFQPOs} the reference energy band has been considered to be 6$-$8 keV to minimise error bars. The observed lags for the lower kHz QPOs and for {LFQPOs} oscillation are on the order of $\sim$ 100$-$200 $\mu s$ and of the order of $\sim$ 8$-$10 milliseconds, generally decreasing with energy, suggesting that softer photons arrive later than harder ones, as expected. In contrast, the energy$-$dependent fractional rms shows an increasing trend with energy, reaching approximately 15$-$18\% at $\sim$ 11 keV across the different states. All energy$-$dependent fractional rms and time lag spectra for the lower kHz QPOs are shown in Figure \ref{Fig06}.  

\subsection{Modelling the Fractional rms and Lag variation}

Having acquired the optimal spectral fitting parameters, we can now focus on identifying the most plausible explanations for the variations in time lag and fractional rms with respect to energy, utilising a straightforward Comptonization model for the lag. Since the observed time lags for lower kHz QPOs are soft, \citet{klitzing:lee2001compton} proposed that a fraction of Comptonized photons could scatter back into the seed photon source. This model has subsequently been employed in other studies \citep{klitzing:kumar2014energy,klitzing:kumar2016constraining,10.1093/mnras/stz3502,2022MNRAS.515.2099B} to explain the energy$-$dependent rms and time lags of kHz QPOs And also of LFQPOs in Black hole binaries \citep{2022MNRAS.515.2099B}. The model\footnote{\url{http://astrosat-ssc.iucaa.in/data_and_analysis}} predicts lag and rms behavior as a function of energy by solving the linearized time$-$dependent Kompaneets equation \citep{klitzing:lee2001compton} in the non$-$relativistic limit ($kT_e << m_e c^2$) accounting for small periodic variations in the seed photon temperature or coronal heating rate considering a fraction, $\eta$, of the Comptonized photons returns to the seed photon source with {an assumption of spherically concentric corona}. The model predicts the time lag and fractional rms variation on the basis of the input spectral parameters like optical depth ($\tau$) {which is related to the spectral photon index via equation \label{equation2} \citep{10.1093/mnras/283.1.193},}

\begin{equation}
\Gamma = \left[\frac{9}{4} + \frac{1}{(kT_e/m_ec^2)\tau\left(1+\tau/3\right)}\right]^{1/2}-\frac{3}{2}
\label{equation2}
\end{equation}

black body temperature, electron temperature, size of the corona and feedback parameter ($\eta$), assuming the input spectrum to the Comptonizing medium to be simple black body in nature. 

\begin{table}
	\centering
    \renewcommand{\arraystretch}{1.5} 
    \setlength{\tabcolsep}{5pt}      

    \begin{tabular}{c c c c c c c c}
		\hline
		\textbf{Obs} & \textbf{Size of} & \textbf{feedback} & $\Delta H$ & $kT_e$ & $kT_{bb}$ & $\tau$ &  {$\nu$}\\
        & \textbf{Corona} & \textbf{Parameter} &  &  &  &  &  \\
		\textbf{No} & (km) & ($\eta$) &  & (keV) & (keV) &  & (Hz)\\
		\hline
		\hline
		1 & 5 & {0.17} & {0.445} & {18.8} & {0.72} & {2.9} & {604}\\
		\hline
		2 & 5 & 0.35 & {0.16} & {4.0} & {0.92} & {6.4} & {733}\\
		\hline
		3 & 5 & 0.32 & {0.195} & {3.15} & {0.75} & {6.4} & {625}\\
		\hline
		4 & 5 & 0.25 & {0.40} & {6.37} & {0.75} & {4.18} & {875}\\
		\hline
	\end{tabular}
	\caption{Details of the theoretical model outcomes that best explain the observed lag and rms variation with respect to energy. Factor represents the feedback coefficient. $\Delta H$, $kT_e$, $kT_{bb}$, $\tau$, and $\nu$ represent heating rate variation, electron temperature, blackbody temperature, optical depth, and frequency at which the lag and rms were calculated.}
	\label{Table6}
\end{table}

In this study, we modelled the spectra using the XSPEC spectral model {\tt{Thcomp*bbody}}, which includes relativistic corrections. For consistency with the timing model, we compared the spectrum derived from the non$-$relativistic equation with the best$-$fit spectrum obtained from {\tt{Thcomp*bbody}}. The values for which the two spectra best fit each other have been taken to calculate the lag and rms. The values are reported in the Table \ref{Table6}. 
There were significant contribution from the disk component that can be observed in the 3$-$5 keV energy regime. To take away the effect of disk emission and reflection, we re$-$scaled the fractional rms data points by $C_T/(C_T-C_{D}-C_{R})$ where $C_T$, $C_{D}$ and $C_{R}$ are the total, disk and reflection count rates in the respective energy bands in different observations since the model itself do not account the contribution from the reflection component.  

The observed lags are of the order of $\sim$ 200$-$300 microseconds, which is larger than the {earlier reported values} of $\sim$ 50$-$100 microseconds, although the uncertainties are large as well. The lag variation can be explained with a corona size of 5 km having a varying feedback factor $\eta$ which ranges between {0.17$-$0.35} from hard to soft. The fractional rms behaviour aligns with the idea of an increasing heating rate variation ranging between {0.16$-$0.445} from soft to hard. This suggests that as the system enters the hard state, the corona intensifies, leading to an increased heating rate and reduced feedback, which in turn results in a smaller soft lag. 

\section{Mass measurement through the Relativistic Precession Model} \label{sec6}

The presence of the twin peak component, along with the broad feature around $\sim$ 30\,Hz, enables us to test the Relativistic Precession Model (RPM) and place constraints on the mass and moment of inertia of the neutron star \citep{1999ApJ...524L..63S,2017MNRAS.465.3581V,2019MNRAS.486.4485D}. 

In RPM model, the $\nu_\phi$, $\nu_{per}$, and $\nu_{nod}$ are identified as Upper kHz, Lower kHz, and { LFQPO} frequency. They are expressed in terms of universal gravitational constant G, speed of the light c, specific angular momentum $\tilde{a} $= $\frac{2\pi I \nu_s}{M}$ , where M is the mass of the NS, $\nu_s$ is the spin frequency of the NS and I is the moment of inertia of the NS. 

\begin{equation}
	\nu_{per} = \nu_{\phi} \left[1-\sqrt{1-\frac{6GM}{c^2r} \pm \frac{8\tilde{a}(GM)^{1/2}}{c^2 r^{3/2}}-\frac{3 \tilde{a}^2}{r^2 c^2}}\right]
	\label{E2}
\end{equation}

\begin{equation}
	\nu_{nod} = \nu_{\phi} \left[1-\sqrt{1 \mp \frac{4\tilde{a}(GM)^{1/2}}{c^2 r^{3/2}}-\frac{3 \tilde{a}^2}{r^2 c^2}}\right]
	\label{E3}
\end{equation}

Here,  r can be expressed in terms of the Upper kHz QPO frequency that is $\nu_\phi$ and can be eliminated from Equation \ref{E2} and \ref{E3} using the following relation.

\begin{equation}
	r = \pm (GM)^{1/3}\left[\frac{1}{2 \pi \nu_\phi} - \frac{\tilde{a}}{c^2}\right]^{2/3}
	\label{E4}
\end{equation}

\begin{figure*}
	\centering
	\begin{tabular}{c c}
		\includegraphics[height=0.35\textheight,width=0.5\textwidth]{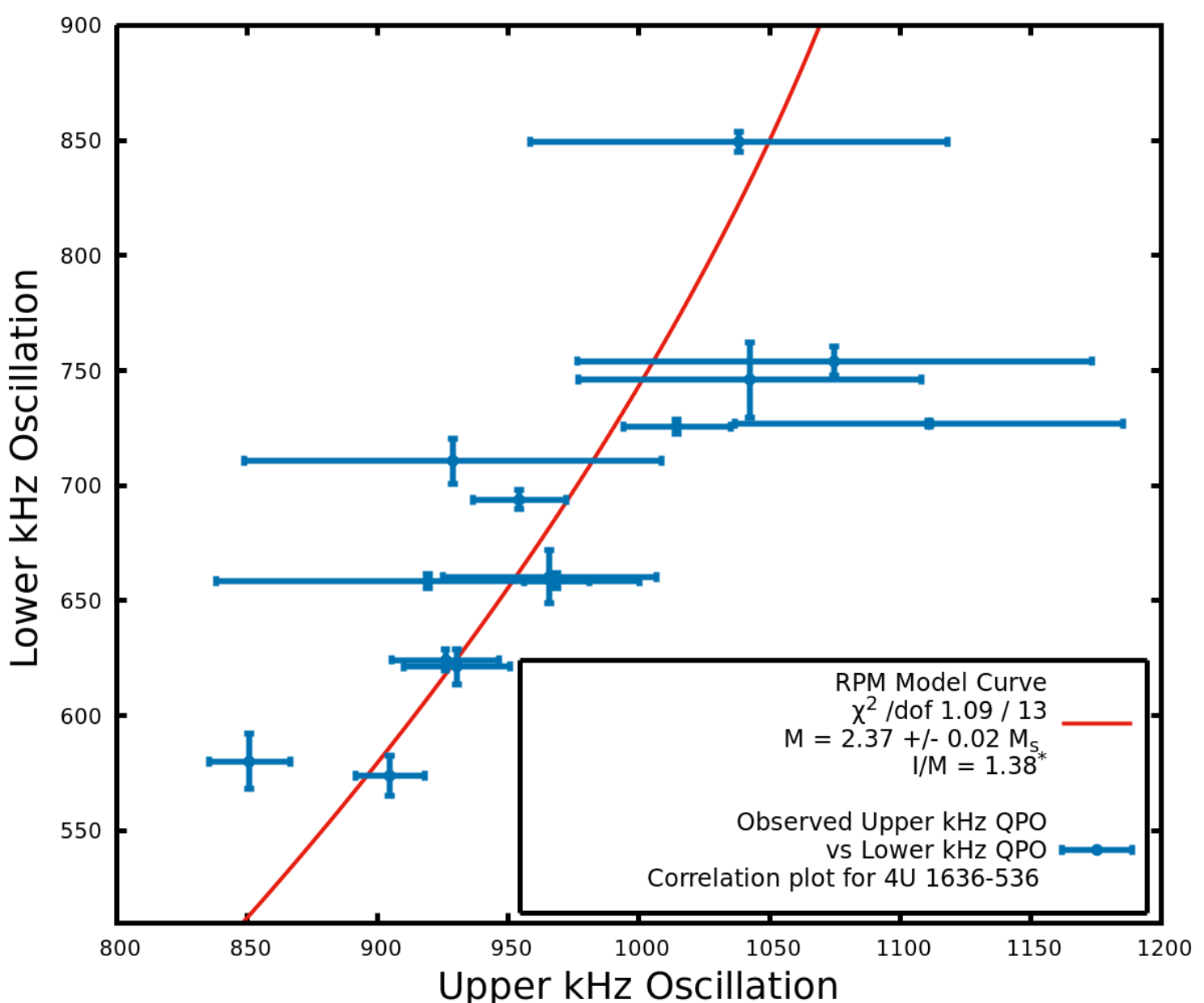} & \includegraphics[height=0.35\textheight,width=0.5\textwidth]{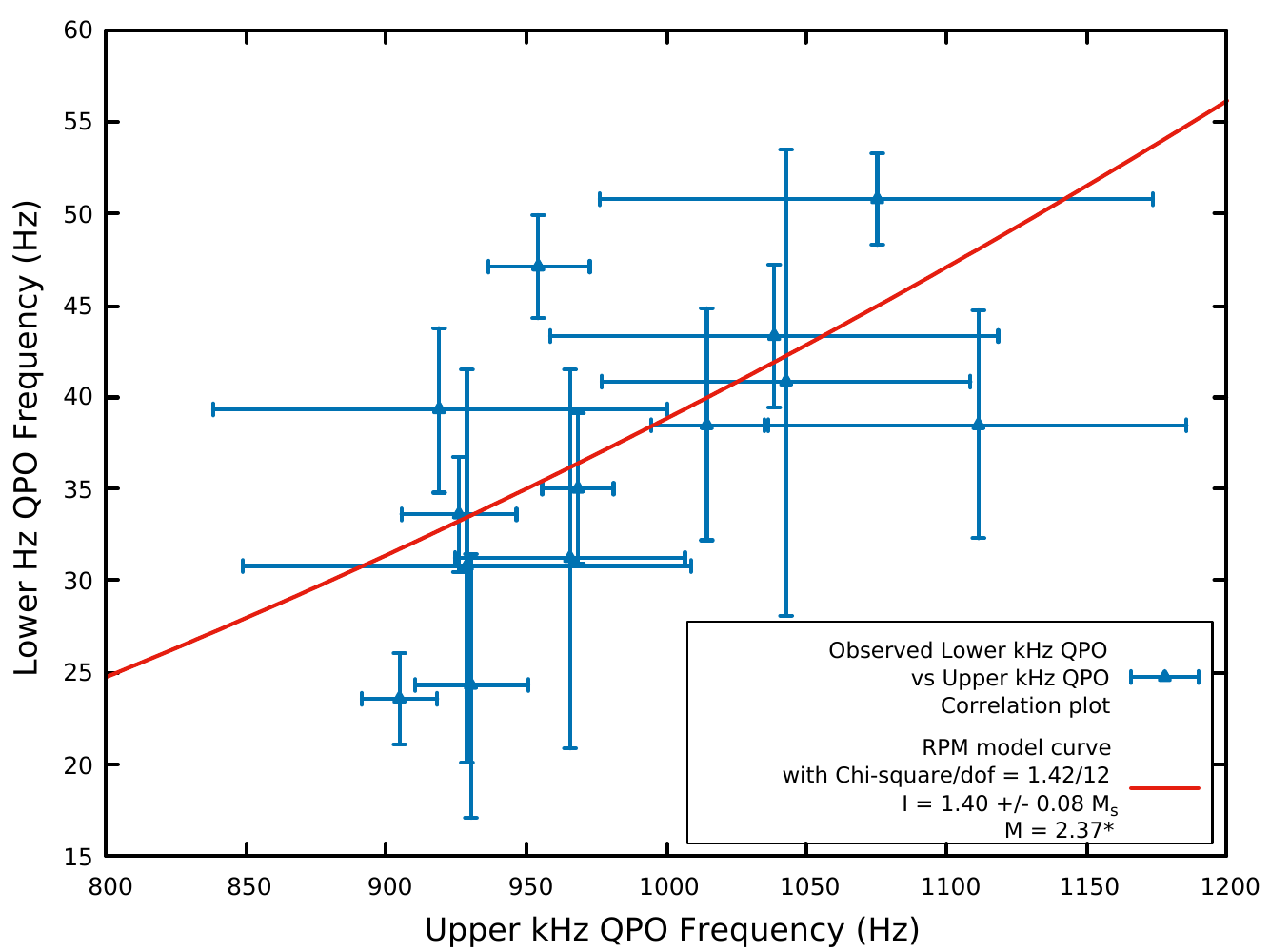} \\
	\end{tabular}
	\caption{In the left panel, the Upper kHz QPO vs lower kHz QPO fitted using RPM model Equation \ref{E2}. In the right panel,  the upper kHz QPO vs { LFQPO} fitted with 2$\times$ Equation \ref{E3}. The estimated mass of the system is 2.37 $\pm$ 0.02 M$_\odot$ and the I/M comes out to be = 1.40 $\pm$ 0.08 I in terms of $I_{45}$.}
	\label{Fig04}
\end{figure*}

Here, positive sign and negative sign represent pro$-$grade and retrograde orbit, respectively.  We have split the Obs 2 and Obs 3 according to the good time intervals and created the PDS in each of the segments to see the evolution of the three frequencies and take the note of these frequencies which are tabulated in the Table \ref{Table4}. Figure \ref{Fig04} (left column) shows the correlation of the upper kHz QPO vs lower kHz QPO and the right column shows the correlation of upper kHz QPO vs { LFQPO} we obtained in different segments. The upper kHz QPO ranges from $\sim$ 850 Hz to $\sim$ 1111 Hz, while lower kHz QPO and the LFQPO are found to vary within $\sim$ 580 Hz to $\sim$ 850 Hz and $\sim$ 23$-$50 Hz. It is to be noted that, in Obs 2 in all the gti segments, we obtained the presence of the three different frequencies, although in Obs 3 we are unable to detect the simultaneous presence of the three frequencies in all the segments. 

Following a similar approach as \citep{Anand_2024}, we have fitted the correlations using the RPM model with Equations \ref{E2} and \ref{E3}. However, we were unable to achieve a satisfactory reduced $\chi^2$ when applying Equation \ref{E3} to the correlation between the upper kHz QPO and the { LFQPO}. To obtain acceptable results, we had to multiply Equation \ref{E3} by a factor of 2, consistent with previous findings \citep{10.1093/mnras/stz2622}, where this factor has been attributed to a two$-$fold symmetry in the system. According to \citep{1999ApL&C..38...57S,Psaltis_1999}, this factor could be an indication to stronger signal production at the even harmonics of the nodal precession frequency due to the tilted disk geometry. From the fit, we have derived the neutron star mass to be 2.37 $\pm$ 0.02 M$_\odot$, and the moment of inertia (I/M$_\odot$) has been found to be 1.40 $\pm$ 0.08 in units of $I_{45}$ \footnote{I in units of 10 $^ {45}$ g cm$^2$ ($I_{45}$).} assuming a spin frequency of $\sim$ 300 Hz. While taking spin frequency 581 Hz the estimated mass of the neutron star comes out to be 2.5 $\pm$ 0.04 M$_\odot$, with (I/M$_\odot$) = 1.35 $\pm$ 0.05. Taking the mass of the NS to be 2.4 M$_\odot$, we can estimate the Schwarzschild radius ($r_s = 2 G M / c^2$) which is $\sim$ 7 km. The inner disk radius is around $6 r_g$ (= 1 ISCO (Inner Most Stable Circular Orbit) ), which is $\sim$ 20 km. {Hence, the inner disc radius we obtained ($\sim$ 20$-$25 km) is greater than the ISCO value seems to be justified.}

\begin{table*}
	\centering
	\renewcommand{\arraystretch}{1.5} 
	\setlength{\tabcolsep}{2pt} 
	\begin{tabular}{c c ccc ccc ccc c}
		\hline
		{\bf Obs} &{\bf Seg} & \multicolumn{3}{c}{\bf Upper kHz QPO} & \multicolumn{3}{c}{\bf Lower kHz QPO} & \multicolumn{3}{c}{\bf { LFQPO}} &  {\bf $\chi^2$/dof}\\
		\hline
		\hline
		& & {$\nu_u$} & {$\sigma_u$ } & {\bf N $\times 10^{-3}$} & {$\nu_L$} & {\bf $\sigma_L$ } & {\bf N $\times 10^{-2}$} & {$\nu_{Lo}$} & {$\sigma_{Lo}$} & {\bf N$\times 10^{-3}$}& \\
		\hline
		2 &	1  & 919.1 $\pm$ 81   & 325.9 $\pm$ 173.9  & 1.17 $\pm$ 0.5  & 658.4 $\pm$ 3.2 & $<$ 33.7    &  3.08 $\pm$ 1.7  & 39.3 $\pm$ 4.5  & 5.80 $\pm$ 5.5    &  1.15 $\pm$ 0.8 &  35.02/30   \\
		&	2  & 928.8 $\pm$ 80 & < 402.2  & 0.19 $\pm$ 0.15  & 710.6 $\pm$ 29.9  &  < 32.6  & 4.9 $\pm$ 1.1 & 30.8 $\pm$ 10.7  & < 32.0   & 2.0 $\pm$ 1.1  &  8.08/10  \\
		&	3  & 965.8 $\pm$ 41  & 150.6 $\pm$ 118.8  & 14.6 $\pm$ 10.2  & 660.3 $\pm$ 11.6 & 84.8 $\pm$ 39.7  &  1.3 $\pm$ 0.5 & 31.2 $\pm$ 10.3    & < 150   & 0.9 $\pm$ 0.4  &  2.20/2     \\
		&   4  & 930.3 $\pm$ 20.2 & 69.4 $\pm$ 68.2 & 4.7 $\pm$ 2.8 &  621.2 $\pm$ 7.76 &  < 52.8 & 0.4 $\pm$ 0.2 & 24.3 $\pm$ 7.2 & < 19.2 & 2.1 $\pm$ 1.3 & 16.27/20 \\
		&   5  & 968.4 $\pm$ 12.5 & 76.5 $\pm$ 40.3 & 5.4 $\pm$ 2.4 & 658.6 $\pm$ 3.4 & 28.4 $\pm$ 20.7 & 0.5 $\pm$ 0.2 &  35.0 $\pm$ 4.1  & 25.8 $\pm$ 14.0 & 3.1 $\pm$ 2.2 & 30.83/40 \\
		&   6  & 971.9 $\pm$ 18.0 & 76.2 $\pm$ 53.1 & 4.4 $\pm$ 1.7 &  693.8 $\pm$ 4.1 & 3.1 $\pm$ 9.3 & 0.6 $\pm$ 0.1 & 18.1 $\pm$ 2.8 & 10.6 $\pm$ 5.4 & 1.4 $\pm$ 0.9 & 45.39/49 \\
		&   7  & 1014.6 $\pm$ 20.4 & 88.0 $\pm$ 84.5 & 2.6 $\pm$ 1.9 &  725.5 $\pm$ 2.9 & 20.1 $\pm$ 11.7 & 0.71 $\pm$ 0.1 & 38.5 $\pm$ 6.3 & 40.0 $\pm$ 26.4 & 2.43 $\pm$ 0.1 & 78.23/72 \\
		&   8  & 1111.1 $\pm$ 74.5 & 220.9 $\pm$ 135.5 & 4.2 $\pm$ 3.7 &  726.9 $\pm$ 1.2 & 13.2 $\pm$ 13.3 & 0.8 $\pm$ 1.2 & 38.5 $\pm$ 6.2 & 45.1 $\pm$ 21.7 & 3.54 $\pm$ 1.23 & 25.35/33 \\
		&   9  & 1074.9 $\pm$ 98.5 & > 68.8 & 7.4 $\pm$ 5.2 &  754.1 $\pm$ 6.4 & 55.3 $\pm$ 20.7 & 0.9 $\pm$ 2.1 & 50.8 $\pm$ 2.5 & 5.1 $\pm$ 4.9 & 0.7 $\pm$ 0.5 & 31.17/42 \\
		\hline
		\hline
		3 & 10  & 936.4 $\pm$ 10.1 & < 192 & 2.9 $\pm$ 1.4 & -- & -- & -- & -- & -- & -- & 33.85/49 \\
		& 11  & 883.9 $\pm$ 8.8 &  65.8 $\pm$ 25.1 & 8.2 $\pm$ 1.8 & -- & -- & -- & -- & -- & -- & 57.68/64 \\
		& 12  & 850.9 $\pm$ 15.5 &  17.5 $\pm$ 11.5 & 7.3 $\pm$ 1.6 & 580.1 $\pm$ 11.9  & < 29.8 & 0.26  $\pm$ 0.05  & --  & -- & -- & 66.78/56 \\
		& 13  & 1038.3 $\pm$ 80 &  < 42.2 & 3.3 $\pm$ 1.6 & 849.4 $\pm$ 4.3  & < 12.3 & 0.7  $\pm$ 0.09  & 43.3 $\pm$ 3.9  & < 18.8  & 0.9  $\pm$ 0.6 & 34.93/51 \\
		& 14  & 1042.3 $\pm$ 65.6 &  < 76.5 & 2.0 $\pm$ 1.5 & 745.9 $\pm$ 16.5  & 118.4 $\pm$  55.2 & 0.9  $\pm$ 0.3  & 40.8 $\pm$ 12.7  & 52.7 $\pm$ 23.3 & 5.8  $\pm$ 2.6 & 88.98/81 \\
		& 15  & 925.9 $\pm$ 20.6 &  235.7 $\pm$ 107.8 & 19.8 $\pm$ 6.7 & 624.2 $\pm$ 4.4  & < 25.3 & 0.4  $\pm$ 0.09  & 33.6 $\pm$ 3.1  & 12.8 $\pm$ 9.5 & 2.3  $\pm$ 1.3 & 107.46/101 \\
		& 16  & 904.7 $\pm$ 13.2 &  111.4 $\pm$ 44.8 & 16.1 $\pm$ 4.3 & 573.8 $\pm$ 8.8  & <80.4 & 0.6 $\pm$ 0.2  & 23.6 $\pm$ 2.5  & 21.8 $\pm$ 8.5 & 6.7  $\pm$ 1.9 & 75.88/86 \\
		& 17  & 903.2 $\pm$ 13.1 &  60.3 $\pm$ 41.7 & 11.4 $\pm$ 5.0 & -- & -- & --  & --  & -- & -- & 64.41/59 \\
		\hline
		
	\end{tabular}
	\caption{Details of the segment and the best fitted parameters for the three frequencies mentioned in the Equation \ref{E2}, and \ref{E3} in Obs 2 and Obs 3. All the errors are reported in 1$\sigma$ confidence level.}
	\label{Table4}
\end{table*}

\section{Summary \& Discussions } \label{sec7}
In this article, we have presented the findings from broadband spectral$-$timing study of 4U 1636$-$536 using four different {\em AstroSat} observations. The HID diagram indicates that the source exhibited different spectral states across observations, showing a progression from a hard to a soft spectral state.

\subsection{Spectral Evolution}

The combined SXT and LAXPC photon spectra, within the 0.7$-$25 keV range across various spectral states of the system, can be modelled by a combination of thermal Comptonized emission with a black body spectrum as input, a relativistic reflection component that takes seed photons from the Comptonized medium, and disk emission. The input blackbody seed photon spectrum is likely due to boundary layer emission from the neutron star surface. The radius of the neutron star is consistent with a typical radius of 10 km in our spectral analysis, considering a distance of 7 kpc.
 
As the state of the system changes, some of the prime spectral parameters change systematically, such as spectral photon index, electron temperature, inner disk temperature and disk normalisation. The spectral parameters noted in Table \ref{Table2} reveal a correlation between black body temperature and inner disk temperature as the system transitions from a harder state (Obs 1) to an extreme soft state (Obs 2), making the spectra more soft component dominated.  
The individual increases in black body ($\sim$14\%) and disk temperature ($\sim$ 40 \%) imply the inward shift of the disk in the soft state, while the decrease in electron temperature suggests a weakening corona as the disk approaches. { This is also evident from the change in the inner disk radius of the system as well. We have observed a significant increase in the reflection component (almost 2 times increase in comparison to the soft state) in the spectra when the system moves towards the hard state.} The ionisation parameter reduces in the hard state (Observation 1). A similar conclusion was previously reached by \citet{1993PASJ...45..605Y} for the NSLMXB 4U 1608-52. This is likely due to a decrease in the solid angle subtended by the reflector as luminosity increases. Such variations suggest changes in the system's geometry, including modifications in the thickness and distance of the reflector, along with other effects such as a reduction in its effective size. Another possible scenario could be the effect of the increased flux from the central compact object. As previously shown by \citet{refId0}, there is a reduction in the ionisation while the blackbody input seed photon flux increases, which is true for this scenario as well.

\subsection{Temporal Evolution}

Alternatively, significant variations are evident in the PDS as the system changes its state. We observe the pronounced presence of the upper kHz QPO, accompanied by a prominent 30$-$40 Hz broad hump feature in the PDS, as the system moves into a softer state, alongside the lower kHz QPO. This evolution enables us to constrain the mass and specific angular momentum of the system using the Relativistic Precession Model (RPM). {Our results yield a mass of $\sim 2.37-2.50 M_\odot$ and an angular momentum of \(1.35-1.40 \, I_{45} / M_\odot\), which is higher than the previously calculated mass of \(1.92 \pm 0.03 \, M_\odot\) with \(I_{45} / M_\odot = 1.07 \pm 0.05\), as obtained for 4U 1728$-$34 by \citet{Anand_2024}. It is to be remembered that although the spins of the two systems may differ but if we take spin to be similar around 300 Hz this still yields a higher mass of the system.} In contrast, \citet{2011ApJ...726...74L}, using alternate models for oscillations, found that the tidal disruption model—which proposes that clumps of orbiting material may expand or compress due to tidal forces, generating oscillation frequencies—better matches the observed data than the RPM. This model better estimates a mass of \(2.40 \pm 0.003 \, M_\odot\) with a spin { (in units of $ J c/GM^2$)} of \(0.17 \pm 0.015\) and \(\chi^2/\text{dof}\) of 18/16 over a mass of \(2.319 \pm 0.005 \, M_\odot\) and {a spin parameter (in units of $ J c/GM^2$)} of \(0.30 \pm 0.001\) with a \(\chi^2/\text{dof}\) of 156/16 under RPM. However, at higher frequencies, the model predicts a sharp decline in the lower kHz QPO, which is not observed in the data they used. This finding also supports and aligns in close agreement to the relatively higher mass that we obtained in our current analysis. The calculated value of ISCO, considering a slowly rotating neutron star found to be less than the colour-corrected inner disk radius of $\sim$ 42 km. In the relatively hard state that is Obs 1, the radius is relatively high $\sim$ 56 km, which is expected.

For the case of the Obs 2 and Obs 3 the presence of the twin kHz QPOs reveals that in both scenarios the {$Q$} of the lower kHz QPOs is almost twice of the upper kHz QPOs {This is in alignment with the previous reports by \citep{10.1111/j.1365-2966.2006.10830.x,10.1111/j.1365-2966.2005.09214.x}}. From the Table \ref{Table4} one can see that the upper kHz QPO frequency can range from $\sim$ 850$-$1111 Hz with a {$Q$} ranging between $\sim$ 5 to $\sim$ 47, and lower kHz QPO frequency can vary from $\sim$ 573$-$850 Hz with a {$Q$} $\sim$ 19$-$70 in these kind of systems. Hence, depending solely on the {$Q$} and centroid frequency, it is difficult to judge whether the single QPO present in Obs 1 and Obs 4 is the upper one or the lower one. Additionally, the absence of the 30 Hz hump in Obs 1 and Obs 4 potentially points to a shift in the system's behaviour, which could serve as an indicator of its state.

\subsection{Finding Radiative Origin}

The findings in this work indicate that for 4U 1636$-$536, the primary X-ray emission originates from a Comptonizing medium surrounding a black body source, likely the neutron star surface. While the estimated inner disk radius comes out to be $\sim$ 20$-$25 km, we obtained a size of the Comptonized medium that can explain the observed lower kHz QPO time$-$lag is consistently to be 5 km. {This indicates that the disk comes very close to the surface of the neutron star system in the soft state. } This is in contradiction to the earlier study on 4U 1702$-$429 by \citet{Chattopadhyay_2024} in the soft state, where the Comptonized medium was found to be relatively larger ($\sim$ 30 km) but still smaller than the estimated inner disk radius of approximately $\sim$ 150 km although, as noted by \citet{Chattopadhyay_2024}, our results supports the idea that the kHz QPO is more likely associated with the Comptonizing medium. In each scenario, as the system transitions from the hard to soft regime, the observed lag and rms variations can be more effectively explained by changes in the variation of the heating rate of the corona. From the Table \ref{Table6}, one can see that the black body temperature and disk temperature increase gradually with the feedback parameter, while the electron temperature decreases. Higher blackbody and disk temperatures imply a more intense radiation from the inner accretion disk and neutron star surface, enhancing the feedback as more photons are available to scatter back. A hotter electron population in the corona typically results in more energetic Comptonized photons due to increased energy transfer during inverse Compton scattering. Higher electron temperatures lead to stronger forward scattering and increased photon escape. This reduces the fraction of photons that return to the disk or neutron star surface. Since the geometry is fixed and fully covering, only the electron temperature plays a crucial role in shaping the fraction of photons that return, thereby reducing the feedback effect as temperature increases; otherwise, it can depend upon the geometry of the corona. Since the feedback effect drives the production of soft lags, a decrease in the feedback coefficient \( \eta \) as the system moves toward the hard state also corresponds to the observed reduction in soft lags within the 3$-$8 keV range in harder spectral regions. 

{ \citet{Vaughan_1998,Kaaret_1999,klitzing:de2013time} had shown that, in lower kHz QPOs, soft photons lag behind hard ones. on the contrary, \citet{klitzing:de2013time} reported that the energy$-$dependent time lags of upper kHz QPOs exhibit a different behaviour, where hard photons lag behind soft ones, producing so$-$called hard lags which led towards a prediction of distinct origins for the lower and upper kHz QPOs.  A subsequent detailed spectral$-$timing analysis of the NS LMXB 4U 1728$-$34 by \citet{klitzing:barret2013soft,Peille_2015} revealed that while RMS increased with energy for both lower and upper kHz QPOs, the time lags of the lower kHz QPOs were consistently soft, whereas those of the upper kHz QPOs were predominantly hard. These authors also fitted the RMS  spectra of 4U 1608$-$52 and 4U 1728$-$34 with a thermal Comptonization model. In addition to it, \citet{klitzing:de2013time} concluded that different physical mechanisms must drive the lower and upper kHz QPOs and that the oscillating spectral component, and thus the QPOs, must be linked to Comptonization. From the energy dependence of the upper kHz QPO lags in 4U 1608$-$52 \citet{Cackett_2016} conclude that the overall shape of the upper kHz QPO lags is qualitatively more consistent with a reverberation origin. A systematic study of multiple sources by \citet{Troyer_2018} further confirmed the differences between upper and lower kHz QPOs. Our analysis also shows a similar soft lag in the lower kHz QPOs as reported earlier. The fractional rms and the time lag spectra can be better explained specifically using the change in the coronal heating rate, which is in agreement with the earlier claims that these features are produced or enhanced in the regions relatively close to the inner accretion flow.} 

\subsection{Limitations}
{
\begin{itemize}
\item We are unable to constrain the upper kHz QPO lags with the given data sets due to its inconsistent presence. This is because the upper-$-$kHz QPO is detected only for a short duration, and its frequency varies across different segments. As a result, using a fixed FWHM and a fixed frequency leads to large error bars. 

\item Moreover, we required a size of corona $\sim $ 40 km to explain the milliseconds lags in the LFQPOs with the given feedback factor in Obs 2 and Obs 3, which indicates a possible different origin of these oscillations, but no such indications have been confirmed from the spectral fitting.

\item Thirdly, the feedback model posits that soft photons from the accretion disk enter the corona, undergo Comptonization, and a fraction of the resulting hard photons is reflected back to the disk, thereby heating it and inducing variability — this cyclical interaction gives rise to the QPO. The inferred coronal size (e.g., $\sim$10 km) represents the characteristic scale of this active feedback region, rather than the full extent of the corona, which may extend further radially and include zones not directly involved in the generation of the kHz QPO.

\item Most importantly, the theoretical understanding of kHz QPOs remains inherently model-dependent. Although our results align with the relativistic precession model (RPM) \citep{1998ApJ...492L..59S, 1999PhRvL..82...17S}, other frameworks—such as the tidal disruption model \citep{2011ApJ...726...74L} and disk oscillation scenarios \citep{kato2007frequency}—can also account for specific features of the observed QPO behavior. Given that no single model fully captures all QPO properties across various sources, it is important to exercise caution when drawing definitive physical interpretations based solely on QPO observations.
\end{itemize}
}

\section{Acknowledments}
In this article, we utilised \textbf{LAXPC} and \textbf{SXT} data from \textit{AstroSat}, provided by the ISSDC (Indian Space Science Data Centre). We appreciate the AstroSat Science Support Cell (ASSC) 's support for scientific analysis. We acknowledge the LAXPC and SXT Payload Operation Centres (POC) at TIFR, Mumbai. We also used software from HEASARC. 
SM express gratitude to the Inter-University Centre for Astronomy and Astrophysics (IUCAA) for the Visiting Associateship Programme, which granted periodic visits to SC and facilitated part of this work. SM sincerely acknowledges financial support from the Department of Space, Government of India, ISRO (grant no. DS\_2B-13013(2)/10/2020-Sec.2). The authors also thank the anonymous referees for their valuable comments and suggestions.


\section{Data Availability}
The LAXPC and SXT archival data that has been used in this article can be found at {\it AstroSat} ISSDC website (\url{ https://astrobrowse.issdc.gov.in/astro\_archive/archive}).

\bibliographystyle{mnras}
\bibliography{paper} 

\begin{thebibliography}{}
\makeatletter
\relax
\def\mn@urlcharsother{\let\do\@makeother \do\$\do\&\do\#\do\^\do\_\do\%\do\~}
\def\mn@doi{\begingroup\mn@urlcharsother \@ifnextchar [ {\mn@doi@}
  {\mn@doi@[]}}
\def\mn@doi@[#1]#2{\def\@tempa{#1}\ifx\@tempa\@empty \href
  {http://dx.doi.org/#2} {doi:#2}\else \href {http://dx.doi.org/#2} {#1}\fi
  \endgroup}
\def\mn@eprint#1#2{\mn@eprint@#1:#2::\@nil}
\def\mn@eprint@arXiv#1{\href {http://arxiv.org/abs/#1} {{\tt arXiv:#1}}}
\def\mn@eprint@dblp#1{\href {http://dblp.uni-trier.de/rec/bibtex/#1.xml}
  {dblp:#1}}
\def\mn@eprint@#1:#2:#3:#4\@nil{\def\@tempa {#1}\def\@tempb {#2}\def\@tempc
  {#3}\ifx \@tempc \@empty \let \@tempc \@tempb \let \@tempb \@tempa \fi \ifx
  \@tempb \@empty \def\@tempb {arXiv}\fi \@ifundefined
  {mn@eprint@\@tempb}{\@tempb:\@tempc}{\expandafter \expandafter \csname
  mn@eprint@\@tempb\endcsname \expandafter{\@tempc}}}

\bibitem[\protect\citeauthoryear{Alpar \& Shaham}{Alpar \&
  Shaham}{1985}]{Alpar1985}
Alpar M.~A.,  Shaham J.,  1985, \mn@doi [Nature] {10.1038/316239a0}, 316, 239

\bibitem[\protect\citeauthoryear{Anand, Misra, Yadav, Jain, Kumar  \&
  Bhattacharya}{Anand et~al.}{2024}]{Anand_2024}
Anand K.,  Misra R.,  Yadav J.~S.,  Jain P.,  Kumar U.,   Bhattacharya D.,
  2024, \mn@doi [The Astrophysical Journal] {10.3847/1538-4357/ad410c}, 967,
  129

\bibitem[\protect\citeauthoryear{Anders \& Grevesse}{Anders \&
  Grevesse}{1989}]{anders1989abundances}
Anders E.,  Grevesse N.,  1989, Geochimica et Cosmochimica acta, 53, 197

\bibitem[\protect\citeauthoryear{{Antia} et~al.,}{{Antia}
  et~al.}{2017}]{klitzing:2017ApJS..231...10A}
{Antia} H.~M.,  et~al., 2017, \mn@doi [\apjs] {10.3847/1538-4365/aa7a0e}, \href
  {https://ui.adsabs.harvard.edu/abs/2017ApJS..231...10A} {231, 10}

\bibitem[\protect\citeauthoryear{Antia et~al.,}{Antia
  et~al.}{2021}]{klitzing:Antia_2021}
Antia H.~M.,  et~al., 2021, \mn@doi [Journal of Astrophysics and Astronomy]
  {10.1007/s12036-021-09712-8}, 42

\bibitem[\protect\citeauthoryear{{Arnaud}}{{Arnaud}}{1996}]{klitzing:1996ASPC..101...17A}
{Arnaud} K.~A.,  1996, in {Jacoby} G.~H.,  {Barnes} J.,  eds,  Astronomical
  Society of the Pacific Conference Series Vol. 101, Astronomical Data Analysis
  Software and Systems V. p.~17

\bibitem[\protect\citeauthoryear{{Bardeen}, {Press}  \& {Teukolsky}}{{Bardeen}
  et~al.}{1972}]{1972ApJ...178..347B}
{Bardeen} J.~M.,  {Press} W.~H.,   {Teukolsky} S.~A.,  1972, \mn@doi [\apj]
  {10.1086/151796}, \href
  {https://ui.adsabs.harvard.edu/abs/1972ApJ...178..347B} {178, 347}

\bibitem[\protect\citeauthoryear{Barret}{Barret}{2013}]{klitzing:barret2013soft}
Barret D.,  2013, \apj, 770, 9

\bibitem[\protect\citeauthoryear{Barret, Olive  \& Miller}{Barret
  et~al.}{2005}]{10.1111/j.1365-2966.2005.09214.x}
Barret D.,  Olive J.-F.,   Miller M.~C.,  2005, \mn@doi [Monthly Notices of the
  Royal Astronomical Society] {10.1111/j.1365-2966.2005.09214.x}, 361, 855

\bibitem[\protect\citeauthoryear{Barret, Olive  \& Miller}{Barret
  et~al.}{2006}]{10.1111/j.1365-2966.2006.10571.x}
Barret D.,  Olive J.-F.,   Miller M.~C.,  2006, \mn@doi [Monthly Notices of the
  Royal Astronomical Society] {10.1111/j.1365-2966.2006.10571.x}, 370, 1140

\bibitem[\protect\citeauthoryear{{Bellavita}, {Garc{\'\i}a}, {M\'endez}  \&
  {Karpouzas}}{{Bellavita} et~al.}{2022}]{2022MNRAS.515.2099B}
{Bellavita} C.,  {Garc{\'\i}a} F.,  {M\'endez} M.,   {Karpouzas} K.,  2022,
  \mn@doi [\mnras] {10.1093/mnras/stac1922}, \href
  {https://ui.adsabs.harvard.edu/abs/2022MNRAS.515.2099B} {515, 2099}

\bibitem[\protect\citeauthoryear{Belloni, Psaltis  \& Van~der Klis}{Belloni
  et~al.}{2002}]{klitzing:belloni2002unified}
Belloni T.,  Psaltis D.,   Van~der Klis M.,  2002, \apj, 572, 392

\bibitem[\protect\citeauthoryear{Belloni, Homan, Motta, Ratti  \&
  Méndez}{Belloni et~al.}{2007}]{10.1111/j.1365-2966.2007.11943.x}
Belloni T.,  Homan J.,  Motta S.,  Ratti E.,   Méndez M.,  2007, \mn@doi
  [Monthly Notices of the Royal Astronomical Society]
  {10.1111/j.1365-2966.2007.11943.x}, 379, 247

\bibitem[\protect\citeauthoryear{{Boutelier}, {Barret}  \&
  {Miller}}{{Boutelier} et~al.}{2009}]{2009MNRAS.399.1901B}
{Boutelier} M.,  {Barret} D.,   {Miller} M.~C.,  2009, \mn@doi [\mnras]
  {10.1111/j.1365-2966.2009.15430.x}, \href
  {https://ui.adsabs.harvard.edu/abs/2009MNRAS.399.1901B} {399, 1901}

\bibitem[\protect\citeauthoryear{Cackett}{Cackett}{2016}]{Cackett_2016}
Cackett E.~M.,  2016, \mn@doi [The Astrophysical Journal]
  {10.3847/0004-637X/826/2/103}, 826, 103

\bibitem[\protect\citeauthoryear{Casares, Cornelisse, Steeghs, Charles, Hynes,
  O'Brien  \& Strohmayer}{Casares
  et~al.}{2006}]{10.1111/j.1365-2966.2006.11106.x}
Casares J.,  Cornelisse R.,  Steeghs D.,  Charles P.~A.,  Hynes R.~I.,  O'Brien
  K.,   Strohmayer T.~E.,  2006, \mn@doi [Monthly Notices of the Royal
  Astronomical Society] {10.1111/j.1365-2966.2006.11106.x}, 373, 1235

\bibitem[\protect\citeauthoryear{Chattopadhyay, Misra, Mandal, Garg  \&
  Pandey}{Chattopadhyay et~al.}{2024}]{Chattopadhyay_2024}
Chattopadhyay S.,  Misra R.,  Mandal S.,  Garg A.,   Pandey S.~K.,  2024,
  \mn@doi [The Astrophysical Journal] {10.3847/1538-4357/ad9332}, 977, 216

\bibitem[\protect\citeauthoryear{{Dauser}, {Wilms}, {Reynolds}  \&
  {Brenneman}}{{Dauser} et~al.}{2010}]{2010MNRAS.409.1534D}
{Dauser} T.,  {Wilms} J.,  {Reynolds} C.~S.,   {Brenneman} L.~W.,  2010,
  \mn@doi [\mnras] {10.1111/j.1365-2966.2010.17393.x}, \href
  {https://ui.adsabs.harvard.edu/abs/2010MNRAS.409.1534D} {409, 1534}

\bibitem[\protect\citeauthoryear{Galloway, Psaltis, Muno  \&
  Chakrabarty}{Galloway et~al.}{2006}]{Galloway_2006}
Galloway D.~K.,  Psaltis D.,  Muno M.~P.,   Chakrabarty D.,  2006, \mn@doi
  [\apj] {10.1086/499579}, 639, 1033

\bibitem[\protect\citeauthoryear{Giles, Hill, Strohmayer  \& Cummings}{Giles
  et~al.}{2002}]{Giles_2002}
Giles A.~B.,  Hill K.~M.,  Strohmayer T.~E.,   Cummings N.,  2002, \mn@doi [The
  Astrophysical Journal] {10.1086/338890}, 568, 279

\bibitem[\protect\citeauthoryear{{Jonker}, {M\'endez}  \& {van der
  Klis}}{{Jonker} et~al.}{2000}]{2000astro.ph..8358J}
{Jonker} P.~G.,  {M\'endez} M.,   {van der Klis} M.,  2000, \mn@doi [arXiv
  e-prints] {10.48550/arXiv.astro-ph/0008358}, \href
  {https://ui.adsabs.harvard.edu/abs/2000astro.ph..8358J} {pp
  astro--ph/0008358}

\bibitem[\protect\citeauthoryear{{Jonker}, {M\'endez}  \& {van der
  Klis}}{{Jonker} et~al.}{2002}]{2002MNRAS.336L...1J}
{Jonker} P.~G.,  {M\'endez} M.,   {van der Klis} M.,  2002, \mn@doi [\mnras]
  {10.1046/j.1365-8711.2002.05781.x}, \href
  {https://ui.adsabs.harvard.edu/abs/2002MNRAS.336L...1J} {336, L1}

\bibitem[\protect\citeauthoryear{{Kaaret}, {Yu}, {Ford}  \& {Zhang}}{{Kaaret}
  et~al.}{1998}]{1998ApJ...497L..93K}
{Kaaret} P.,  {Yu} W.,  {Ford} E.~C.,   {Zhang} S.~N.,  1998, \mn@doi [\apjl]
  {10.1086/311279}, \href
  {https://ui.adsabs.harvard.edu/abs/1998ApJ...497L..93K} {497, L93}

\bibitem[\protect\citeauthoryear{Kaaret, Piraino, Ford  \& Santangelo}{Kaaret
  et~al.}{1999}]{Kaaret_1999}
Kaaret P.,  Piraino S.,  Ford E.~C.,   Santangelo A.,  1999, \mn@doi [The
  Astrophysical Journal] {10.1086/311941}, 514, L31

\bibitem[\protect\citeauthoryear{Karpouzas, Méndez, Ribeiro, Altamirano, Blaes
   \& García}{Karpouzas et~al.}{2020}]{10.1093/mnras/stz3502}
Karpouzas K.,  Méndez M.,  Ribeiro E.~M.,  Altamirano D.,  Blaes O.,   García
  F.,  2020, \mn@doi [Monthly Notices of the Royal Astronomical Society]
  {10.1093/mnras/stz3502}, 492, 1399

\bibitem[\protect\citeauthoryear{Kato}{Kato}{2007}]{kato2007frequency}
Kato S.,  2007, Publications of the Astronomical Society of Japan, 59, 451

\bibitem[\protect\citeauthoryear{{Kolehmainen}, {Done}  \& {D{\'\i}az
  Trigo}}{{Kolehmainen} et~al.}{2011}]{2011MNRAS.416..311K}
{Kolehmainen} M.,  {Done} C.,   {D{\'\i}az Trigo} M.,  2011, \mn@doi [\mnras]
  {10.1111/j.1365-2966.2011.19040.x}, \href
  {https://ui.adsabs.harvard.edu/abs/2011MNRAS.416..311K} {416, 311}

\bibitem[\protect\citeauthoryear{Kumar \& Misra}{Kumar \&
  Misra}{2014}]{klitzing:kumar2014energy}
Kumar N.,  Misra R.,  2014, \mnras, 445, 2818

\bibitem[\protect\citeauthoryear{Kumar \& Misra}{Kumar \&
  Misra}{2016}]{klitzing:kumar2016constraining}
Kumar N.,  Misra R.,  2016, \mnras, 461, 2580

\bibitem[\protect\citeauthoryear{{Lamb} \& {Miller}}{{Lamb} \&
  {Miller}}{2003}]{2003astro.ph..8179L}
{Lamb} F.~K.,  {Miller} M.~C.,  2003, \mn@doi [arXiv e-prints]
  {10.48550/arXiv.astro-ph/0308179}, \href
  {https://ui.adsabs.harvard.edu/abs/2003astro.ph..8179L} {pp
  astro--ph/0308179}

\bibitem[\protect\citeauthoryear{Lee \& Miller}{Lee \&
  Miller}{1998}]{klitzing:lee1998comptonization}
Lee H.~C.,  Miller G.~S.,  1998, \mnras, 299, 479

\bibitem[\protect\citeauthoryear{Lee, Misra  \& Taam}{Lee
  et~al.}{2001}]{klitzing:lee2001compton}
Lee H.~C.,  Misra R.,   Taam R.~E.,  2001, \apj, 549, L229

\bibitem[\protect\citeauthoryear{{Lin}, {Boutelier}, {Barret}  \&
  {Zhang}}{{Lin} et~al.}{2011}]{2011ApJ...726...74L}
{Lin} Y.-F.,  {Boutelier} M.,  {Barret} D.,   {Zhang} S.-N.,  2011, \mn@doi
  [\apj] {10.1088/0004-637X/726/2/74}, \href
  {https://ui.adsabs.harvard.edu/abs/2011ApJ...726...74L} {726, 74}

\bibitem[\protect\citeauthoryear{{Lyu, M.}, {Zhang, G. B.}, {Wang, H. G.}  \&
  {García, F.}}{{Lyu, M.} et~al.}{2023}]{refId0}
{Lyu, M.} {Zhang, G. B.} {Wang, H. G.}  {García, F.} 2023, \mn@doi [A&A]
  {10.1051/0004-6361/202346584}, 677, A156

\bibitem[\protect\citeauthoryear{Mancuso et~al.,}{Mancuso
  et~al.}{2023}]{10.1093/mnras/stad949}
Mancuso G.~C.,  et~al., 2023, \mn@doi [Monthly Notices of the Royal
  Astronomical Society] {10.1093/mnras/stad949}, 521, 5616

\bibitem[\protect\citeauthoryear{{Mastichiadis}, {Petropoulou}  \&
  {Kylafis}}{{Mastichiadis} et~al.}{2022}]{2022A&A...662A.118M}
{Mastichiadis} A.,  {Petropoulou} M.,   {Kylafis} N.~D.,  2022, \mn@doi [\aap]
  {10.1051/0004-6361/202243397}, \href
  {https://ui.adsabs.harvard.edu/abs/2022A&A...662A.118M} {662, A118}

\bibitem[\protect\citeauthoryear{M\'endez \& Belloni}{M\'endez \&
  Belloni}{2020}]{klitzing:M_ndez_2020}
M\'endez M.,  Belloni T.~M.,  2020, in , Timing Neutron Stars: Pulsations,
  Oscillations and Explosions.
Springer Berlin Heidelberg, pp 263--331, \mn@doi{10.1007/978-3-662-62110-3_6},
  \url {https://doi.org/10.1007%2F978-3-662-62110-3_6}

\bibitem[\protect\citeauthoryear{{M\'endez}, {van der Klis}, {van Paradijs},
  {Lewin}, {Lamb}, {Vaughan}, {Kuulkers}  \& {Psaltis}}{{M\'endez}
  et~al.}{1997}]{1997ApJ...485L..37M}
{M\'endez} M.,  {van der Klis} M.,  {van Paradijs} J.,  {Lewin} W.~H.~G.,
  {Lamb} F.~K.,  {Vaughan} B.~A.,  {Kuulkers} E.,   {Psaltis} D.,  1997,
  \mn@doi [\apjl] {10.1086/310803}, \href
  {https://ui.adsabs.harvard.edu/abs/1997ApJ...485L..37M} {485, L37}

\bibitem[\protect\citeauthoryear{{M\'endez}, {van der Klis}, {Wijnands},
  {Ford}, {van Paradijs}  \& {Vaughan}}{{M\'endez}
  et~al.}{1998}]{1998ApJ...505L..23M}
{M\'endez} M.,  {van der Klis} M.,  {Wijnands} R.,  {Ford} E.~C.,  {van
  Paradijs} J.,   {Vaughan} B.~A.,  1998, \mn@doi [\apjl] {10.1086/311600},
  \href {https://ui.adsabs.harvard.edu/abs/1998ApJ...505L..23M} {505, L23}

\bibitem[\protect\citeauthoryear{Merloni, Vietri, Stella  \& Bini}{Merloni
  et~al.}{1999}]{10.1046/j.1365-8711.1999.02307.x}
Merloni A.,  Vietri M.,  Stella L.,   Bini D.,  1999, \mn@doi [Monthly Notices
  of the Royal Astronomical Society] {10.1046/j.1365-8711.1999.02307.x}, 304,
  155

\bibitem[\protect\citeauthoryear{{Miller}}{{Miller}}{1999}]{1999ApJ...515L..77M}
{Miller} M.~C.,  1999, \mn@doi [\apjl] {10.1086/311970}, \href
  {https://ui.adsabs.harvard.edu/abs/1999ApJ...515L..77M} {515, L77}

\bibitem[\protect\citeauthoryear{{Miller}, {Lamb}  \& {Psaltis}}{{Miller}
  et~al.}{1998}]{1998ApJ...508..791M}
{Miller} M.~C.,  {Lamb} F.~K.,   {Psaltis} D.,  1998, \mn@doi [\apj]
  {10.1086/306408}, \href
  {https://ui.adsabs.harvard.edu/abs/1998ApJ...508..791M} {508, 791}

\bibitem[\protect\citeauthoryear{{Mitsuda} et~al.,}{{Mitsuda}
  et~al.}{1984}]{1984PASJ...36..741M}
{Mitsuda} K.,  et~al., 1984, \pasj, \href
  {https://ui.adsabs.harvard.edu/abs/1984PASJ...36..741M} {36, 741}

\bibitem[\protect\citeauthoryear{{Motta}, {Rouco Escorial}, {Kuulkers},
  {Mu{\~n}oz-Darias}  \& {Sanna}}{{Motta} et~al.}{2017}]{2017MNRAS.468.2311M}
{Motta} S.~E.,  {Rouco Escorial} A.,  {Kuulkers} E.,  {Mu{\~n}oz-Darias} T.,
  {Sanna} A.,  2017, \mn@doi [\mnras] {10.1093/mnras/stx570}, \href
  {https://ui.adsabs.harvard.edu/abs/2017MNRAS.468.2311M} {468, 2311}

\bibitem[\protect\citeauthoryear{Méndez}{Méndez}{2006}]{10.1111/j.1365-2966.2006.10830.x}
Méndez M.,  2006, \mn@doi [Monthly Notices of the Royal Astronomical Society]
  {10.1111/j.1365-2966.2006.10830.x}, 371, 1925

\bibitem[\protect\citeauthoryear{Nowak}{Nowak}{2000}]{klitzing:nowak2000there}
Nowak M.~A.,  2000, \mnras, 318, 361

\bibitem[\protect\citeauthoryear{Peille, Barret  \& Uttley}{Peille
  et~al.}{2015a}]{klitzing:peille2015spectral}
Peille P.,  Barret D.,   Uttley P.,  2015a, \apj, 811, 109

\bibitem[\protect\citeauthoryear{Peille, Barret  \& Uttley}{Peille
  et~al.}{2015b}]{Peille_2015}
Peille P.,  Barret D.,   Uttley P.,  2015b, \mn@doi [The Astrophysical Journal]
  {10.1088/0004-637X/811/2/109}, 811, 109

\bibitem[\protect\citeauthoryear{Peirano \& Méndez}{Peirano \&
  Méndez}{2021}]{klitzing:10.1093/mnras/stab1905}
Peirano V.,  Méndez M.,  2021, \mn@doi [\mnras] {10.1093/mnras/stab1905}, 506,
  2746

\bibitem[\protect\citeauthoryear{Peirano \& Méndez}{Peirano \&
  Méndez}{2022}]{10.1093/mnras/stac1071}
Peirano V.,  Méndez M.,  2022, \mn@doi [Monthly Notices of the Royal
  Astronomical Society] {10.1093/mnras/stac1071}, 513, 2804

\bibitem[\protect\citeauthoryear{Psaltis, Belloni  \& van~der Klis}{Psaltis
  et~al.}{1999}]{Psaltis_1999}
Psaltis D.,  Belloni T.,   van~der Klis M.,  1999, \mn@doi [The Astrophysical
  Journal] {10.1086/307436}, 520, 262

\bibitem[\protect\citeauthoryear{Ribeiro, Méndez, Zhang  \& Sanna}{Ribeiro
  et~al.}{2017}]{10.1093/mnras/stx1686}
Ribeiro E.~M.,  Méndez M.,  Zhang G.,   Sanna A.,  2017, \mn@doi [Monthly
  Notices of the Royal Astronomical Society] {10.1093/mnras/stx1686}, 471, 1208

\bibitem[\protect\citeauthoryear{Ribeiro, Méndez, de Avellar, Zhang  \&
  Karpouzas}{Ribeiro et~al.}{2019}]{10.1093/mnras/stz2463}
Ribeiro E.~M.,  Méndez M.,  de Avellar M. G.~B.,  Zhang G.,   Karpouzas K.,
  2019, \mn@doi [Monthly Notices of the Royal Astronomical Society]
  {10.1093/mnras/stz2463}, 489, 4980

\bibitem[\protect\citeauthoryear{{Roy}, {Beri}  \& {Bhattacharyya}}{{Roy}
  et~al.}{2021}]{2021MNRAS.508.2123R}
{Roy} P.,  {Beri} A.,   {Bhattacharyya} S.,  2021, \mn@doi [\mnras]
  {10.1093/mnras/stab2680}, \href
  {https://ui.adsabs.harvard.edu/abs/2021MNRAS.508.2123R} {508, 2123}

\bibitem[\protect\citeauthoryear{{Schnerr}, {Reerink}, {van der Klis}, {Homan},
  {M{\'e}ndez}, {Fender}  \& {Kuulkers}}{{Schnerr}
  et~al.}{2003}]{2003A&A...406..221S}
{Schnerr} R.~S.,  {Reerink} T.,  {van der Klis} M.,  {Homan} J.,  {M{\'e}ndez}
  M.,  {Fender} R.~P.,   {Kuulkers} E.,  2003, \mn@doi [\aap]
  {10.1051/0004-6361:20030682}, \href
  {https://ui.adsabs.harvard.edu/abs/2003A&A...406..221S} {406, 221}

\bibitem[\protect\citeauthoryear{{Stella} \& {Vietri}}{{Stella} \&
  {Vietri}}{1998}]{1998ApJ...492L..59S}
{Stella} L.,  {Vietri} M.,  1998, \mn@doi [\apjl] {10.1086/311075}, \href
  {https://ui.adsabs.harvard.edu/abs/1998ApJ...492L..59S} {492, L59}

\bibitem[\protect\citeauthoryear{{Stella} \& {Vietri}}{{Stella} \&
  {Vietri}}{1999}]{1999PhRvL..82...17S}
{Stella} L.,  {Vietri} M.,  1999, \mn@doi [\prl] {10.1103/PhysRevLett.82.17},
  \href {https://ui.adsabs.harvard.edu/abs/1999PhRvL..82...17S} {82, 17}

\bibitem[\protect\citeauthoryear{{Stella}, {Vietri}  \& {Morsink}}{{Stella}
  et~al.}{1999a}]{1999ApL&C..38...57S}
{Stella} L.,  {Vietri} M.,   {Morsink} S.,  1999a, Astrophysical Letters and
  Communications, \href {https://ui.adsabs.harvard.edu/abs/1999ApL&C..38...57S}
  {38, 57}

\bibitem[\protect\citeauthoryear{{Stella}, {Vietri}  \& {Morsink}}{{Stella}
  et~al.}{1999b}]{1999ApJ...524L..63S}
{Stella} L.,  {Vietri} M.,   {Morsink} S.~M.,  1999b, \mn@doi [\apjl]
  {10.1086/312291}, \href
  {https://ui.adsabs.harvard.edu/abs/1999ApJ...524L..63S} {524, L63}

\bibitem[\protect\citeauthoryear{{Strohmayer}, {Zhang}, {Swank}, {Smale},
  {Titarchuk}, {Day}  \& {Lee}}{{Strohmayer}
  et~al.}{1996}]{klitzing:1996ApJ...469L...9S}
{Strohmayer} T.~E.,  {Zhang} W.,  {Swank} J.~H.,  {Smale} A.,  {Titarchuk} L.,
  {Day} C.,   {Lee} U.,  1996, \mn@doi [\apjl] {10.1086/310261}, \href
  {https://ui.adsabs.harvard.edu/abs/1996ApJ...469L...9S} {469, L9}

\bibitem[\protect\citeauthoryear{Strohmayer, Zhang, Swank, White  \&
  Lapidus}{Strohmayer et~al.}{1998}]{Strohmayer_1998}
Strohmayer T.~E.,  Zhang W.,  Swank J.~H.,  White N.~E.,   Lapidus I.,  1998,
  \mn@doi [The Astrophysical Journal] {10.1086/311322}, 498, L135

\bibitem[\protect\citeauthoryear{Titarchuk, Lapidus  \& Muslimov}{Titarchuk
  et~al.}{1998}]{Titarchuk_1998}
Titarchuk L.,  Lapidus I.,   Muslimov A.,  1998, \mn@doi [The Astrophysical
  Journal] {10.1086/305642}, 499, 315

\bibitem[\protect\citeauthoryear{{Troyer}, {Cackett}, {Peille}  \&
  {Barret}}{{Troyer} et~al.}{2018a}]{klitzing:2018ApJ...860..167T}
{Troyer} J.~S.,  {Cackett} E.~M.,  {Peille} P.,   {Barret} D.,  2018a, \mn@doi
  [\apj] {10.3847/1538-4357/aac4a4}, \href
  {https://ui.adsabs.harvard.edu/abs/2018ApJ...860..167T} {860, 167}

\bibitem[\protect\citeauthoryear{Troyer, Cackett, Peille  \& Barret}{Troyer
  et~al.}{2018b}]{Troyer_2018}
Troyer J.~S.,  Cackett E.~M.,  Peille P.,   Barret D.,  2018b, \mn@doi [The
  Astrophysical Journal] {10.3847/1538-4357/aac4a4}, 860, 167

\bibitem[\protect\citeauthoryear{Tse, Galloway, Chou, Heger  \& Hsieh}{Tse
  et~al.}{2020}]{10.1093/mnras/staa3224}
Tse K.,  Galloway D.~K.,  Chou Y.,  Heger A.,   Hsieh H.-E.,  2020, \mn@doi
  [Monthly Notices of the Royal Astronomical Society] {10.1093/mnras/staa3224},
  500, 34

\bibitem[\protect\citeauthoryear{Vaughan et~al.,}{Vaughan
  et~al.}{1998}]{Vaughan_1998}
Vaughan B.~A.,  et~al., 1998, \mn@doi [The Astrophysical Journal]
  {10.1086/311785}, 509, L145

\bibitem[\protect\citeauthoryear{{Verner}, {Ferland}, {Korista}  \&
  {Yakovlev}}{{Verner} et~al.}{1996}]{1996ApJ...465..487V}
{Verner} D.~A.,  {Ferland} G.~J.,  {Korista} K.~T.,   {Yakovlev} D.~G.,  1996,
  \mn@doi [\apj] {10.1086/177435}, \href
  {https://ui.adsabs.harvard.edu/abs/1996ApJ...465..487V} {465, 487}

\bibitem[\protect\citeauthoryear{{Wang}, {Chen}, {Zhang}, {Lei}  \&
  {Qu}}{{Wang} et~al.}{2014}]{2014AN....335..168W}
{Wang} D.~H.,  {Chen} L.,  {Zhang} C.~M.,  {Lei} Y.~J.,   {Qu} J.~L.,  2014,
  \mn@doi [Astronomische Nachrichten] {10.1002/asna.201311995}, \href
  {https://ui.adsabs.harvard.edu/abs/2014AN....335..168W} {335, 168}

\bibitem[\protect\citeauthoryear{Wang et~al.}{Wang
  et~al.}{2016}]{klitzing:wang2016brief}
Wang J.,  et~al., 2016, International Journal of Astronomy and Astrophysics, 6,
  82

\bibitem[\protect\citeauthoryear{Wang, Méndez, Sanna, Altamirano  \&
  Belloni}{Wang et~al.}{2017}]{10.1093/mnras/stx671}
Wang Y.,  Méndez M.,  Sanna A.,  Altamirano D.,   Belloni T.~M.,  2017,
  \mn@doi [Monthly Notices of the Royal Astronomical Society]
  {10.1093/mnras/stx671}, 468, 2256

\bibitem[\protect\citeauthoryear{Wijers, Van~Paradijs  \& Lewin}{Wijers
  et~al.}{1987}]{klitzing:wijers1987energy}
Wijers R.,  Van~Paradijs J.,   Lewin W.,  1987, \mnras, 228, 17P

\bibitem[\protect\citeauthoryear{Wijnands, Méndez, van~der Klis, Psaltis,
  Kuulkers  \& Lamb}{Wijnands et~al.}{1998}]{Wijnands_1998}
Wijnands R.,  Méndez M.,  van~der Klis M.,  Psaltis D.,  Kuulkers E.,   Lamb
  F.~K.,  1998, \mn@doi [The Astrophysical Journal] {10.1086/311564}, 504, L35

\bibitem[\protect\citeauthoryear{Willmore, Mason, Sanford, Hawkins, Murdin,
  Penston  \& Penston}{Willmore et~al.}{1974}]{10.1093/mnras/169.1.7}
Willmore A.~P.,  Mason K.~O.,  Sanford P.~W.,  Hawkins F.~J.,  Murdin P.,
  Penston M.~V.,   Penston M.~J.,  1974, \mn@doi [Monthly Notices of the Royal
  Astronomical Society] {10.1093/mnras/169.1.7}, 169, 7

\bibitem[\protect\citeauthoryear{{Wilms}, {Allen}  \& {McCray}}{{Wilms}
  et~al.}{2000}]{klitzing:2000ApJ...542..914W}
{Wilms} J.,  {Allen} A.,   {McCray} R.,  2000, \mn@doi [\apj] {10.1086/317016},
  \href {https://ui.adsabs.harvard.edu/abs/2000ApJ...542..914W} {542, 914}

\bibitem[\protect\citeauthoryear{{Yoshida}, {Mitsuda}, {Ebisawa}, {Ueda},
  {Fujimoto}, {Yaqoob}  \& {Done}}{{Yoshida}
  et~al.}{1993}]{1993PASJ...45..605Y}
{Yoshida} K.,  {Mitsuda} K.,  {Ebisawa} K.,  {Ueda} Y.,  {Fujimoto} R.,
  {Yaqoob} T.,   {Done} C.,  1993, \pasj, \href
  {https://ui.adsabs.harvard.edu/abs/1993PASJ...45..605Y} {45, 605}

\bibitem[\protect\citeauthoryear{Zdziarski, Johnson  \& Magdziarz}{Zdziarski
  et~al.}{1996}]{10.1093/mnras/283.1.193}
Zdziarski A.~A.,  Johnson W.~N.,   Magdziarz P.,  1996, \mn@doi [Monthly
  Notices of the Royal Astronomical Society] {10.1093/mnras/283.1.193}, 283,
  193

\bibitem[\protect\citeauthoryear{{Zdziarski}, {Szanecki}, {Poutanen},
  {Gierli{\'n}ski}  \& {Biernacki}}{{Zdziarski}
  et~al.}{2020}]{2020MNRAS.492.5234Z}
{Zdziarski} A.~A.,  {Szanecki} M.,  {Poutanen} J.,  {Gierli{\'n}ski} M.,
  {Biernacki} P.,  2020, \mn@doi [\mnras] {10.1093/mnras/staa159}, \href
  {https://ui.adsabs.harvard.edu/abs/2020MNRAS.492.5234Z} {492, 5234}

\bibitem[\protect\citeauthoryear{{Zhang}}{{Zhang}}{2004}]{2004A&A...423..401Z}
{Zhang} C.,  2004, \mn@doi [\aap] {10.1051/0004-6361:20035808}, \href
  {https://ui.adsabs.harvard.edu/abs/2004A&A...423..401Z} {423, 401}

\bibitem[\protect\citeauthoryear{{Zhang}, {Lapidus}, {White}  \&
  {Titarchuk}}{{Zhang} et~al.}{1996}]{1996ApJ...473L.135Z}
{Zhang} W.,  {Lapidus} I.,  {White} N.~E.,   {Titarchuk} L.,  1996, \mn@doi
  [\apjl] {10.1086/310411}, \href
  {https://ui.adsabs.harvard.edu/abs/1996ApJ...473L.135Z} {473, L135}

\bibitem[\protect\citeauthoryear{Zhang, Strohmayer  \& Swank}{Zhang
  et~al.}{1997}]{Zhang_1997}
Zhang W.,  Strohmayer T.~E.,   Swank J.~H.,  1997, \mn@doi [The Astrophysical
  Journal] {10.1086/310719}, 482, L167–L170

\bibitem[\protect\citeauthoryear{{Zhang}, {Yin}, {Zhao}, {Zhang}  \&
  {Song}}{{Zhang} et~al.}{2006}]{2006MNRAS.366.1373Z}
{Zhang} C.~M.,  {Yin} H.~X.,  {Zhao} Y.~H.,  {Zhang} F.,   {Song} L.~M.,  2006,
  \mn@doi [\mnras] {10.1111/j.1365-2966.2006.09920.x}, \href
  {https://ui.adsabs.harvard.edu/abs/2006MNRAS.366.1373Z} {366, 1373}

\bibitem[\protect\citeauthoryear{de Avellar, M\'endez, Sanna  \&
  Horvath}{de~Avellar et~al.}{2013}]{klitzing:de2013time}
de Avellar M.~G.,  M\'endez M.,  Sanna A.,   Horvath J.~E.,  2013, \mnras, 433,
  3453

\bibitem[\protect\citeauthoryear{{du Buisson}, {Motta}  \& {Fender}}{{du
  Buisson} et~al.}{2019}]{2019MNRAS.486.4485D}
{du Buisson} L.,  {Motta} S.,   {Fender} R.,  2019, \mn@doi [\mnras]
  {10.1093/mnras/stz1160}, \href
  {https://ui.adsabs.harvard.edu/abs/2019MNRAS.486.4485D} {486, 4485}

\bibitem[\protect\citeauthoryear{{van Doesburgh} \& {van der Klis}}{{van
  Doesburgh} \& {van der Klis}}{2017}]{2017MNRAS.465.3581V}
{van Doesburgh} M.,  {van der Klis} M.,  2017, \mn@doi [\mnras]
  {10.1093/mnras/stw2961}, \href
  {https://ui.adsabs.harvard.edu/abs/2017MNRAS.465.3581V} {465, 3581}

\bibitem[\protect\citeauthoryear{{van der Klis}}{{van der
  Klis}}{1997}]{VanDerKlis_1997}
{van der Klis} M.,  1997, in , Astronomical Time Series.
Springer Netherlands, pp 121--132, \mn@doi{10.1007/978-94-015-8941-3_10}, \url
  {https://doi.org/10.1007%2F978-94-015-8941-3_10}

\bibitem[\protect\citeauthoryear{van~der Klis}{van~der
  Klis}{2006a}]{vanderKlis_2006}
van~der Klis M.,  2006a, Rapid X-ray variability.
Cambridge Astrophysics, Cambridge University Press

\bibitem[\protect\citeauthoryear{{van der Klis}}{{van der
  Klis}}{2006b}]{2006csxs.book...39V}
{van der Klis} M.,  2006b, in {Lewin} W. H.~G.,  {van der Klis} M.,  eds, ,
  Vol.~39, Compact stellar X-ray sources.
pp 39--112

\bibitem[\protect\citeauthoryear{{van der Klis}, {Swank}, {Zhang}, {Jahoda},
  {Morgan}, {Lewin}, {Vaughan}  \& {van Paradijs}}{{van der Klis}
  et~al.}{1996}]{1996ApJ...469L...1V}
{van der Klis} M.,  {Swank} J.~H.,  {Zhang} W.,  {Jahoda} K.,  {Morgan} E.~H.,
  {Lewin} W.~H.~G.,  {Vaughan} B.,   {van Paradijs} J.,  1996, \mn@doi [\apjl]
  {10.1086/310251}, \href
  {https://ui.adsabs.harvard.edu/abs/1996ApJ...469L...1V} {469, L1}

\bibitem[\protect\citeauthoryear{{van der Klis}, {Wijnands}, {Horne}  \&
  {Chen}}{{van der Klis} et~al.}{1997}]{1997ApJ...481L..97V}
{van der Klis} M.,  {Wijnands} R. A.~D.,  {Horne} K.,   {Chen} W.,  1997,
  \mn@doi [\apjl] {10.1086/310656}, \href
  {https://ui.adsabs.harvard.edu/abs/1997ApJ...481L..97V} {481, L97}

\bibitem[\protect\citeauthoryear{van Doesburgh \&
  van der Klis}{van Doesburgh \&
  van der Klis}{2019}]{10.1093/mnras/stz2622}
van Doesburgh M.,  van der Klis M.,  2019, \mn@doi [Monthly Notices of the
  Royal Astronomical Society] {10.1093/mnras/stz2622}, 490, 5270

\makeatother
\end{thebibliography}

\bsp	
\label{lastpage}
\end{document}